\documentclass[onecollarge,natbib,referee]{svjour2}
\bibpunct{[}{]}{;}{n}{}{,} 
\smartqed  
\usepackage{amsmath,revsymb,graphicx,amssymb,youngtab}

\newcommand{\kt}[1]{\ensuremath{|{#1}\rangle}}
\newcommand{\br}[1]{\ensuremath{\langle {#1}|}}
\newcommand{\bk}[2]{\ensuremath{\langle {#1}|{#2}\rangle}}
\newcommand{\ykt}[1]{\ensuremath{\left| \Yvcentermath1 {#1} \Yvcentermath0 \right\rangle}}
\newcommand{\ybr}[1]{\ensuremath{\left\langle\Yvcentermath1 {#1} \Yvcentermath0 \right|}}
\newcommand{\rykt}[1]{\ensuremath{\left\| \Yvcentermath1 {#1} \Yvcentermath0 \right\rangle}}
\newcommand{\rybr}[1]{\ensuremath{\left\langle\Yvcentermath1 {#1} \Yvcentermath0 \right\|}}
\newcommand{\ybk}[2]{\ensuremath{\left\langle\Yvcentermath1 {#1} \Yvcentermath0 \right|\left.\Yvcentermath1 {#2} \Yvcentermath0 \right\rangle}}
\newcommand{\oybk}[2]{\ensuremath{\left\langle\Yvcentermath1 {#1} \Yvcentermath0 \Large| \Yvcentermath1 {#2} \Yvcentermath0 \right\rangle}}
\newcommand{\HS}{\mathcal{H}}
\newcommand{\SHS}{\mathcal{S}}
\newcommand{\KHS}{\mathcal{K}}
\newcommand{\pp}[1]{\ensuremath{{\lfloor #1 \rfloor }}}

\journalname{Few-Body Systems}
\begin{document}

\title{One-Dimensional Traps, Two-Body Interactions, Few-Body Symmetries}

\subtitle{I.~One, Two, and Three Particles}

\titlerunning{One, Two, Few: I.~1, 2, 3}        

\author{N.L.\ Harshman}


\institute{N.L.\ Harshman \at
              Department of Physics\\
              American University\\
              4400 Massachusetts Ave. NW\\
              Washington, DC 20016-8058 USA\\
              Tel.: +1-202-885-3479\\
              Fax: +1-202-885-2723\\
              \email{harshman@american.edu}           
}

\date{Received: date / Accepted: date}

\maketitle

\begin{abstract}
This is the first in a pair of articles that classify the configuration space and kinematic symmetry groups for $N$ identical particles in one-dimensional traps experiencing Galilean-invariant two-body interactions. These symmetries explain degeneracies in the few-body spectrum and demonstrate how tuning the trap shape and the particle interactions can manipulate these degeneracies. The additional symmetries that emerge in the non-interacting limit and in the unitary limit of an infinitely strong contact interaction are sufficient to algebraically solve for the spectrum and degeneracy in terms of the one-particle observables. Symmetry also determines the degree to which the algebraic expressions for energy level shifts by weak interactions or nearly-unitary interactions are universal, i.e.\ independent of trap shape and details of the interaction. Identical fermions and bosons with and without spin are considered. This article sequentially analyzes the symmetries of one, two and three particles in asymmetric, symmetric, and harmonic traps; the sequel article treats the $N$ particle case. 
\keywords{One-dimensional traps \and Few-body symmetries \and Unitary limit of contact interaction}
\end{abstract}

\section{Introduction to Part I}
\label{intro}

The focus of this pair of articles is the non-relativistic, one-dimensional, few-body Hamiltonian with the following characteristics: (1) Each particle has the same mass and experiences the same trapping potential. 
(2) There is a two-body interaction term for each pair that depends only on the distance between particles.
(3) Each particle has a finite number of internal levels that do not participate directly in the trap or two-body interactions.  The particles could be distinguishable, or they could be identical bosons or fermions. The total Hamiltonian for the system can be expressed as
\begin{subequations}\label{model}
\begin{equation}
\hat{H}^N = \sum_{i=1}^N \hat{H}^1_i + \sum_{i<j}^N \hat{V}_{ij}.
\end{equation}
Denoting each canonical pair of particle observables by $[\hat{Q}_j,\hat{P}_k]= i \delta_{jk}$ and choosing natural units, the one-body Hamiltonian for particle $i$ is
\begin{equation}\label{oneH}
\hat{H}^1_i = \frac{1}{2} \hat{P}_i{}^2 + V^1(\hat{Q}_i).
\end{equation}
The two-body interaction term $\hat{V}_{ij}$ has the Galilean invariance property $\hat{V}_{ij}=V^2(|\hat{Q}_i-\hat{Q}_j|)$. Particular attention is focused on the contact interaction, expressed in particle coordinates ${ \bf q}=(q_1,q_2,\ldots,q_N)$ as
\begin{equation}\label{contact}
V^2(|q_i-q_j|) = g \delta(q_i - q_j).
\end{equation}
\end{subequations}

The goal of this pair of articles is to classify the symmetries of the few-body Hamiltonian $\hat{H}^N$ for the cases of no interaction, general interaction, and unitary limit of contact interaction and then to demonstrate how these symmetries can be used to calculate spectral properties and understand universal features. Two classes of symmetries are considered: \emph{configuration space symmetries} and \emph{kinematic symmetries}. By configuration space symmetry, I mean the group of transformations of configuration space $\mathcal{Q}^N\sim\mathbb{R}^N$ that are represented as unitary operators that commute with $\hat{H}^N$. Configuration space symmetry includes the permutation group of identical particles, but it also can include parity or emergent symmetries depending on the trap $V^1$ and interaction $V^2$ potentials. Kinematic symmetry is realized by the group of \emph{all} unitary operators that commute with $\hat{H}^N$. The kinematic symmetry group necessarily contains the configuration space symmetry and time translation as subgroups. A key insight is that the dimensions of the irreducible representations of the kinematic symmetry group (if properly identified) explain the degeneracies in the spectrum of $\hat{H}^N$.

This first article analyzes the symmetries of one, two, and three particles. The configuration space symmetries and kinematics symmetries are developed incrementally, and the ways in which the trap shape and the two-body interaction effect the symmetry are explained with examples. For systems with few degrees of freedom, the order of finite symmetry groups are small, so explicit calculations and applications are included. Additionally, the symmetries of one, two and three particles can be visualized using familiar geometrical methods and analogies. The sequel article treats the general case of $N$ particles. In that case, the formal, algebraic machinery of group representation theory demonstrates its power. However, the price is a higher degree of abstraction and the necessity of computer-based algebraic methods.

\subsection{Motivation}

The model Hamiltonian (\ref{model}) has a long history inspired by applications to atomic, molecular, nuclear and condensed matter physics. Going back to the beginnings of quantum mechanics, various subfields have given different names (e.g.\ Stoner Hamiltonian, Tonks-Girardeau gas, Lieb-Liniger model, no-core shell model) to particular instances of the model and its higher dimensional generalizations. There is a large mathematical physics literature on the one-dimensional model, and certain cases of $\hat{H}^N$ are exemplars of solvability in few-body and many-body systems~\cite{Albeverio, sutherland, Cazalilla2011, Guan2013, Gaudin}. The increasingly precise preparation, control and measurement of ultracold trapped atomic systems in effectively one-dimensional traps~\cite{Olshanii1998} is driving another surge of theoretical interest in this few-body model, c.f.\ \cite{Girardeau2004, Oelkers2006, Girardeau2007, Deuretzbacher2008, Hao2009, Guan2009, Yang2009, Ma2009, Guan2010, Girardeau2010a, Girardeau2010b, Girardeau2011, Fang2011, Brouzos2012a, Koscik2012, Armstrong2012, Harshman2012, Brouzos2012b, Gharashi2013, Bugnion2013, Sowinski2013, Volosniev2013, Wilson2014, Deuretzbacher2014, DAmico2014, Cui2014, Lindgren2014, Harshman2014, Garcia2014, Volosniev2014, Levinsen2014, Loft2014, Yang2015}. Few body properties can drive the dynamics of many-body cold atom trapped systems, like trap loss and equilibration, and few-body observables may be more directly accessed by tunneling rates and spectroscopic methods.

\begin{figure}
\centering
\includegraphics[width=\linewidth]{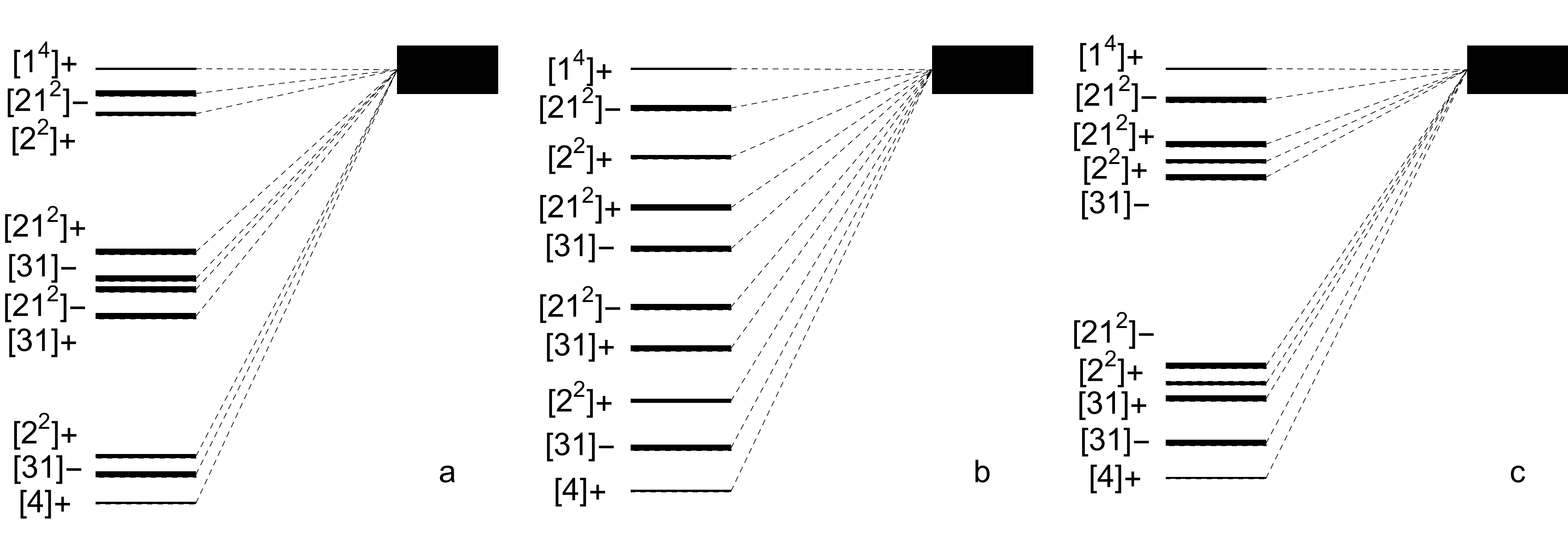}
\caption{Level splitting diagram for four particles in three symmetric traps with contact interactions: (a) double well; (b) infinite square well; (c) V-shaped or cusped well, i.e. softer than harmonic. The thick band on the right in each figure is the 24-fold degenerate ground state energy level for four distinguishable particles in the unitary limit $g\rightarrow \infty$ of the contact interaction. This level carries two representations of $\mathrm{S}_4$ simultaneously; one copy corresponds to particle permutation symmetry $\mathrm{P}_4$ and the other to ordering permutation symmetry. In the near unitary limit, ordering permutation symmetry $\mathfrak{O}_4$ is broken by tunneling and the energy levels split into irreducible representations of $\mathrm{P}_4$ with either even or odd parity. The thickness of the line indicates the degeneracy of these levels for distinguishable identical particles without spin. For example, the irrep labeled $[4]^+$ is the non-degenerate, positive parity, totally symmetric spatial state. It can be occupied by bosons with any number of internal levels, or fermions with at least four internal levels.  The three trap shapes are distinguished by the ratio of tunneling amplitudes $t/u$, where $t$ is the tunneling amplitude for the left-most or right-most particle to exchange with the adjacent inner particle and $u$ is the tunneling amplitude for the two inner particles to exchange. To first order, both are proportional to $1/g$. The following ratios have been chosen to illustrate the trap dependence of these amplitudes: (a) $t/u=2.9$; (b) $t/u = 1$; (c) $t/u=0.3$. The idea is that (a) for double wells tunneling in the middle is suppressed so $t>u$; (b) for infinite square well the potential is uniform inside the trap and so (for low particle density) $t$ and $u$ are approximately the same; (c) for softer wells, there is more phase space in the middle of the well so $u>t$. For harmonic wells, $t/u \approx 0.762 $ (c.f.\ \cite{Deuretzbacher2014,Volosniev2013,Levinsen2014,Gharashi2015}). In subfigure (c) a more extreme ratio is depicted, corresponding to a V-shaped or cusped trap.}
\label{fig:level4}
\end{figure}

Although group theory has a long history of being productive in quantum mechanics, the ``Gruppenpest''\footnote{See the Introduction to \cite{Chen} for a discussion of the Gruppenpest.} can be so frustrating that it is customary to begin with an explanation of why all this mathematical apparatus is worth the effort. The essential claim is that the symmetry classifications provided in this article can be exploited for qualitative, analytic and numerical studies of few-body systems trapped in one dimension and they provide a unifying framework for this recent wave of analysis. These methods solve or simplify numerous questions about the spectrum, degeneracy and dynamics, including the following:
\begin{itemize}
\item identical particle symmetrization,
\item perturbation theory from the non-interacting to the weak interaction limit,
\item perturbation theory from the unitary limit of the contact interaction to the nearly-unitary limit,
\item methods of exact diagonalization in truncated Hilbert spaces,
\item perturbation theory for not-quite identical particles,
\item adiabatic or non-adiabatic particle dynamics under variation of interaction parameters or trap shape, and
\item trial wave functions for variational or Monte Carlo methods.
\end{itemize}
As a preview of the kind of results that symmetry classification and calculations provide, see Fig.\ \ref{fig:level4}. It depicts how level splitting in the near-unitary limit of the contact interaction depends on trap shape for four particles. Depending on whether the particles are fermions or bosons, with or without spin, only certain energy levels can be populated. The method of calculation is developed later in the paper, but the main idea is that near unitarity, level splitting is determined by the tunneling amplitudes of adjacent particles and these tunneling amplitudes depend on the shape of the trap. The energy eigenstates can be found by diagonalizing a tunneling operator, and these eigenstates carry irreducible representations of the symmetric group for four particles $\mathrm{S}_4$ and for the parity symmetry.

Many applications of the representation theory of the symmetric group already exist in the recent literature; a few examples relevant to these articles are \cite{Guan2009, Yang2009, Ma2009, Fang2011, Harshman2012, Cui2014, Harshman2014}. Parity is also widely exploited, and the special symmetries of harmonic traps are often explicitly or implicitly invoked. The focus of this article is to see how much more solvability is provided by additional configuration space and kinematic symmetries inherited from the trap shape and the Galilean invariance of the interactions. We know that in the case of the infinite square well and contact interactions of any strength, there is enough symmetry to provide integrability, i.e.\ the Bethe ansatz solutions (c.f.\ \cite{Gaudin, Oelkers2006}). The experimental tunability of few-body symmetries and the close connection between finite groups and integrability~\cite{Chanu,Batchelor2014} suggest novel possibilities for embodying mathematical structures in ultracold atomic systems.

Symmetry also aids the study of ``universal'' few body phenomena, a term used (with some local variation) to describe dynamical effects that do not depend strongly on the particular details of the constituent few body systems or on the nature of their interactions. See \cite{Cazalilla2011, Gangardt2004,Farrell2010} for discussions of universality in one-dimension. Universal properties established in atomic systems can also reveal themselves in few-body systems at the chemical or nuclear scale. Universality can drive the dynamics of coherence, entanglement and equilibration in certain many-body systems. One approach to universality is to figure out how much about the few-body system can be inferred from the symmetries of $\hat{H}^N$ without specific knowledge of the trap or the interaction. The relationships among trap shape, interaction, and permutation symmetry determine which  properties of the system can be algebraically solved for in terms of one-particle observables. The degree to which a few-body system possesses this kind of `algebraic solvability' is at least some component of universality. The best example is provided by the unitary limit of the contact interaction, which has enough symmetry to be exactly solved for any $N$ given the one-particle spectrum~\cite{Girardeau2007, Deuretzbacher2008, Guan2009, Yang2009, Ma2009, Girardeau2010a, Girardeau2010b, Girardeau2011, Fang2011, Cui2014, Harshman2014}. The question of how level splitting in the weak interaction limit and near-unitary limit depends on trap shape is a theme that runs throughout this pair of articles.

Symmetry methods also provide geometrical insight into the highly-abstract interplay of trap shape, interaction, spin, and particle symmetrization. Especially for low particle numbers, symmetries can be pictured and manipulated in the mind. To a large extent, the geometrical constructions and geometrical methods applied in the works \cite{Yang2009, Brouzos2012a, Gharashi2013, Volosniev2013, Garcia2014, Loft2014, Bao2013} motivated this article.   In this first article, I argue that by analyzing the cases of two and three particles geometrically, we get insights that can guide us for higher particle numbers where more abstract methods are required.

\subsection{Outline of the Articles}

The next section of this article explains which configuration space and kinematic symmetries are possible for one particle in asymmetric and symmetric traps, and explains the extra kinematic symmetry that occurs for the harmonic trap. The third and fourth sections develop symmetry classifications and techniques for two and three particles. In each scenario, the non-interacting case is considered first, then the interacting case (including weak interactions), and finally the unitary limit of the contact interaction (including the near-unitary limit). For three particles, state permutation symmetry and ordering permutation symmetry are introduced as useful concepts complementary to the more familiar particle permutation symmetry. Along the way, a variety of applications, diagrams and figures are included that attempt to make the symmetry methods more concrete and less abstract. This article ends with a conclusion that reflects on what this symmetry analysis says about universality in this model.

The second article in this series derives the general form of the symmetry classifications for $N$ particles. It is necessarily more technical (and has fewer pictures). After an introduction that gives the expressions for the minimal configuration space and kinematic symmetries inherited by the construction of the few-body system from the one-body systems with two-body interactions, the next section gives an overview of the symmetric group $\mathrm{S}_N$ and its representations. Definitions, notation and conventions necessary to extend these methods for $N>3$ are briefly reviewed. In particular, a kind of $\mathrm{S}_N$ representation space called a permutation module is shown to be especially useful for the analysis of $N$ identical particles. The third section establishes the symmetries for $N$ non-interacting particles and describes the geometric realization of particle permutations and other symmetries in configuration space. The irreducible representations for the minimal kinematic symmetry group are derived and state permutation symmetry is used to construct a complete set of commuting observables that facilitates identical particle symmetrization. A final result of this section establishes the isomorphism between the bosonic non-interacting spectrum and the fermionic spectrum (which remains invariant under contact interactions). The fourth section classifies the symmetries for $N$ particles interacting via two-body Galilean invariant potentials. The symmetries of two-body matrix elements are derived and state permutation symmetry makes another appearance, this time as a property of the contact interaction. The two body matrix elements are used to analyze level splitting in the weak interaction limit and I conjecture that algebraic solvability is lost for more than five multicomponent particles. The fifth section treats the unitary limit of the contact interaction. Ordering permutation symmetry emerges as new symmetry of the system, and the near-unitary limit can be understood in terms of symmetry breaking of ordering permutation symmetry by tunneling among different sectors of configuration space. The final and concluding section of both articles summarizes how the main results relate to the question of universality and describes some possible further extensions and applications of this work.

\subsection{A Few Notes about Group Notation}

These articles are addressed to several distinct audiences, including novices and experts, interested in low-dimensional, trapped ultracold atomic systems, general quantum few-body systems, and/or mathematical physics. I have attempted to clearly signpost the content into sections, subsections and subsubsections so that readers can pick and choose what matches their interests and background. The first article is more pedagogical and less technical. The second article presumes more familiarity with group representation theory, but most necessary ideas are developed in this first article.

Another challenge when providing clarity for a diverse audience is in the choice of notation. This is particularly important when taking about symmetry, groups, and group representations because different physical symmetries may be isomorphic to the same abstract group, and groups usually have multiple inequivalent representations. The next few subsubsections provide a brief introduction to the notations for symmetries, abstract groups, and their representations that will be used in these articles.

\subsubsection{Groups and Representations}

The configuration space symmetry group for an $N$ particle system with Hamiltonian (\ref{model}) is denoted $\mathrm{C}_N$. For the particular case when there are no two-particle interactions, the configuration space symmetry is denoted $\mathrm{C}_N^0$ and for the unitary limit of the contact interaction the group is denoted $\mathrm{C}_N^\infty$. The kinematic symmetries are similarly denoted $\mathrm{K}^N$, $\mathrm{K}_N^0$, and $\mathrm{K}_N^\infty$. 

These symmetry groups are isomorphic to abstract groups. The specific abstract group depends on the shape of the trap. For example, for non-interacting particles in an harmonic trap, $\mathrm{K}_N^0$ is isomorphic to $\mathrm{U}(N)$, the abstract group realized by unitary $N\times N$ matrices. To highlight the distinction between isomorphic and equality, I write $\mathrm{K}_N^0 \sim \mathrm{U}(N)$.

For a given group $\mathrm{G}$, up to three different representations are in the analysis of this article:
\begin{enumerate}
\item Unitary irreducible representations, or irreps. Almost all groups in this article are finite or compact, and they have a finite or countable number of finite-dimensional irreps. Other representations are built out of direct sums of irreps. The labels or notations for irreps depend on the group. As an example, pretend the symbol $\spadesuit$ labels a particular irrep of the group $\mathrm{G}$. The dimension of the irrep is $d(\mathrm{G};\spadesuit)$, or  $d(\spadesuit)$ if the group is obvious from the irrep label. The $d(\spadesuit)$-dimensional unitary matrix representation of $g\in\mathrm{G}$ is denoted $\underline{D}^\spadesuit(g)$. The  complex vector space that carries the representation $\underline{D}^\spadesuit$ is $\mathcal{M}^\spadesuit \sim \mathbb{C}^{d(\spadesuit)}$.

\item Hilbert space representation. Every element of a symmetry group $g\in\mathrm{G}$ is represented by a unitary operator on the Hilbert space, denoted $\hat{U}(g)$ or $\hat{g}$. The Hilbert space 
can be decomposed into irreps of the symmetry group. For example, if the irreps of $\mathrm{G}$ are $\spadesuit$, $\heartsuit$, $\diamondsuit$, and $\clubsuit$, then
\[
\HS = \HS^\spadesuit \oplus \HS^\heartsuit \oplus \HS^\diamondsuit \oplus \HS^\clubsuit.
\]
Each subspace of $\HS$ could be a single irrep, i.e.\ $\HS^\spadesuit \sim \mathcal{M}^\spadesuit$, but generally the subspace  $\HS^\spadesuit$ is a tower of irrep spaces
\[
\HS^\spadesuit = \bigoplus_i \HS^{\spadesuit}_i 
\]
where $i$ is a label or set of labels that distinguish different copies $\HS^{\spadesuit}_i \sim \mathcal{M}^\spadesuit$ of equivalent irreps  of $\mathrm{G}$ in $\HS^\spadesuit$.

\item Configuration space representation. This refers to the action of $\mathrm{C}_N$ (or the special cases $\mathrm{C}_N^0$ or $\mathrm{C}_N^\infty$) on the configuration space $(q_1,q_2,\ldots,q_N)\in \mathcal{Q}^N=\mathbb{R}^N$. The representation of $g\in\mathrm{C}_N$ is denoted $\underline{O}(g)$. Typically, this representation of $\mathrm{C}_N$ is not irreducible. 
\end{enumerate}

\subsubsection{Symmetric Group}

One obvious symmetry of the model Hamiltonian (\ref{model}) is the group of particle permutations $\mathrm{P}_N$.  The configuration space symmetries $\mathrm{C}_N$, $\mathrm{C}^0_N$, and $\mathrm{C}^\infty_N$ all must contain $\mathrm{P}_N$ as a subgroup. The elements $p\in\mathrm{P}_N$ can be described either in permutation notation or cycle notation. For example, the same permutation $p$ can be written either as permutation $\{312\}$ or three-cycle $(132)$. Both notations for $p$ describe the map in which particle 1 is replaced by particle 3, particle 2 is replaced by 1, and particle 3 is replaced by 2.

The group $\mathrm{P}_N$ is isomorphic to the abstract group $\mathrm{S}_N$, the symmetric group on $N$ objects. Two other groups described in later sections are also isomorphic to symmetric groups, the group of state permutations $\mathfrak{P}_\pp{\nu}$ on the state composition $\pp{\nu}$ and the group of ordering permutations $\mathfrak{O}_N$ on $N$ particles. The symmetric group $\mathrm{S}_N$ has order $N!$ and it has irreps labeled by whole number partitions of $N$ denoted $[\mu]=[\mu_1 \mu_2 \cdots \mu_r]$ where $\sum_i \mu_i = N$. These irreps are sometimes depicted by Ferrers diagrams (also called Young diagrams), which are $r$ rows of boxes with $\mu_r$ boxes in each row. The $\mathrm{S}_N$ irrep spaces are denoted $\mathcal{M}^{[\mu]}$ and the matrix  representation of $p\in\mathrm{S}_N$ on $\mathcal{M}^{[\mu]}$ is denoted $\underline{D}^{[\mu]}(p)$. 

A few notes and examples with $\mathrm{S}_N$ for $N=1$, $2$, and $3$:
\begin{enumerate}
\item The group $\mathrm{S}_1 \sim \mathrm{Z}_1$ is trivial.
\item The group $\mathrm{S}_2 \sim \mathrm{Z}_2$ has two elements $e$ and $(12)$ and it is abelian. It has two one-dimensional irreps labeled by the Ferrers diagrams $\tiny \yng(2)$ and $\tiny\yng(1,1)$, or more compactly by partitions $[2]$ and $[11]\equiv [1^2]$. For the trivial, symmetric representation, we have $\underline{D}^{[2]}(12) = 1$ and for the faithful, antisymmetric representation $\underline{D}^{[1^2]}(12) = -1$.
\item The group $\mathrm{S}_3$ has six elements in three classes: the identity $e$, three two-cycles (or transpositions) $(12)$, $(23)$, and $(31)$, and two three-cycles $(123)$ and $(132)$. This group is not abelian, and since it is not abelian, no faithful representations can be one-dimensional. There are three irreps: $[3]$ or $\tiny\yng(3)$ (one-dimensional, symmetric); $[21]$ or $\tiny\yng(2,1)$ (two-dimensional, faithful); and $[111]\equiv [1^3]$ or $\tiny\yng(1,1,1)$ (one-dimensional, antisymmetric).
\end{enumerate}

\subsubsection{Point Groups}

Most of  the configuration space symmetries considered in this article are point groups. Point groups are orthogonal transformations of the $N$-particle configuration space $\mathcal{Q}^N=\mathbb{R}^N$, and the set of all possible point groups for a given dimension is completely characterized~\cite{Coxeter}. When many people think of symmetry, it is the geometrical realization of point-group invariant objects that they envision. One important class of point groups are the finite Coxeter groups. These are generated by reflections in Euclidean space, they are the symmetries of regular polyhedra, and their categorization is closely related to the structure of simple Lie algebras.

The maximal point group for $\mathcal{Q}^N$ is the group $\mathrm{O}(N)$ of all orthogonal transformations in $N$ dimensions, i.e.\ all reflections and rotations. All other point groups in $N$ dimensions are subgroups of $\mathrm{O}(N)$. Here are a few facts about point groups in low dimensions useful for understanding this article:
\begin{enumerate}
\item In one dimension, there are only two point groups. One is the trivial group that contains just the identity $e$. The other is the group $\mathrm{O}(1)$ that contains the identity $e$ and a single reflection $\Pi$. These groups are isomorphic to the abstract cyclic groups $\mathrm{Z}_1$ and $\mathrm{Z}_2$, respectively. Since these groups are both abelian, all irreps are one dimensional. Irreps of $\mathrm{O}(1)$ are labeled by $\pi=\pm 1$.
\item In two dimensions, besides $\mathrm{O}(2)$, there are two series of finite-order point groups. This article employs several of the dihedral groups $\mathrm{D}_k$, finite groups of order $2k$ that include $k$ rotations (including the identity) and $k$ reflections. The group $\mathrm{D}_1\sim\mathrm{Z}_2$ is the symmetry of a butterfly, the group $\mathrm{D}_2$ is the symmetry of a rectangle, the group $\mathrm{D}_4$ is the symmetry of a square. For $k>2$, the group $\mathrm{D}_k$ is not abelian and so its faithful irreps are not one-dimensional.
\item In three dimensions, besides $\mathrm{O}(3)$, there are seven infinite series of finite-order point groups, seven other finite-order point groups, and four other continuous point groups. These groups are familiar to some from chemical or solid states physics; see \cite{Hamermesh} for a palatable introduction to these groups and their irreps. There are multiple conventions for the notation of three-dimensional groups and irreps (Sch\"onflies, Coxeter, orbifold, etc.); specific notations are introduced as necessary.
\end{enumerate}

\subsubsection{A Few Other Groups and Notes}

The abstract group of translation by a single parameter $x\in\mathbb{R}$ is denoted $\mathrm{T}_x$. Irreps of $\mathrm{T}_x$ are one-dimensional and abelian and characterized by a single real number. Specific examples include: the group of time translation $\mathrm{T}_t$ represented on the Hilbert space by $\hat{U}(t)=\exp(-i\hat{H}t)$ with irrep labels called energy; and the group of space translations $\mathrm{T}_a$ represented by $\hat{U}(a)=\exp(-i\hat{P}a)$ with irrep labels called momentum.

Finally, note the following:
\begin{itemize}
\item Groups are generally denoted by capital Roman letters, e.g.\ $\mathrm{S}_N$, $\mathrm{C}_N$, $\mathrm{U}(N)$, etc.
\item Vector spaces are denoted by capital calligraphic letters, e.g.\ the total Hilbert space $\mathcal{H}$, the spatial Hilbert space $\mathcal{K}$, the spin (or internal component) Hilbert space $\mathcal{S}$, irrep spaces $\mathcal{M}$, or the configuration space $\mathcal{Q}$.
\item Operators on the infinite-dimensional Hilbert space have `hats' like $\hat{Q}_i$ and $\hat{R}_i$. Their eigenvalues are usually lowercase like $q_i$ and $r_i$. Matrix operators on finite-dimensional spaces like irrep spaces $\mathcal{M}$ or configuration space $\mathcal{Q}$ are underlined, e.g.\ the representations $\underline{D}$ and $\underline{O}$.
\item Ordered sequences of numbers or symbols are denoted by angle brackets, e.g.\ $\langle \nu \rangle  = \langle 3, 0, 2, 4, 4, 0 \rangle$ and $\langle \mu \rangle =  \langle \alpha, \beta, \alpha \rangle$. Compositions of unordered numbers or symbols are denoted by floor brackets. For example, the compositions of the previous two sequences are $\pp{\nu}  = \pp{002344}= \pp{0^2234^2}$ and $\pp{\mu} = \pp{\alpha\alpha\beta} =\pp{\alpha^2\beta}$. The set of all sequences with a composition $\pp{\nu}$ is the same as all permutations  of the composition, denoted $\mathrm{P}_\pp{\nu}$. The shape of a composition is the pattern of degeneracies in a composition, e.g.\ $[\nu]=[2211] = [2^21^2]$ and $[\mu]=[21]$, and always corresponds to a partition of the length of the sequence.
\item The non-negative integers $\{0,1,2,3,4,\cdots\}$ are denoted $\mathbb{N}$.
\end{itemize}

\section{One-Particle Symmetries}

This section describes  the configuration space symmetry group $\mathrm{C}_1$ and the kinematic symmetry group $\mathrm{K}_1$ for one-particle in asymmetric, symmetric and harmonic traps. The symmetries $\mathrm{C}_1$ and $\mathrm{K}_1$ are built from basic abstract groups that have only one-dimensional representations. These one-particle symmetry groups are the building blocks of the multi-particle analysis.

Consider one particle in a one-dimensional trap and denote its spatial Hilbert space $\KHS$. The total Hilbert space $\HS=\KHS \otimes \SHS$ is the tensor product of the spatial Hilbert space and the spin Hilbert space (discussed at the end of this section). All one-dimensional systems have at least the symmetry group $\mathrm{T}_t$ of time translations. Although this observation seems trivial, this symmetry is enough to guarantee integrability for any one-dimensional system. The abelian, one-parameter group of time translations $\mathrm{T}_t$ has one-dimensional irreps labeled by the energy $\epsilon$ and the set of allowed energies determined by the Hamiltonian $\hat{H}^1$ is the spectrum $\sigma_1$. Time translation group is represented by exponentiation of the Hamiltonian $\hat{U}(t) = \exp(-i \hat{H}^1 t)$. 

For a single particle trapped in one dimension, the energy spectrum $\sigma_1=\{\epsilon_0,\epsilon_1,\ldots\}$  is discrete, countably-infinite and non-degenerate.  An energy spectrum with this simple form excludes the important idealized case of infinite lattices and periodic boundary conditions. Further, a discrete spectrum is only a low-energy approximation for wells with finite depth because it does not have a continuous piece. There are probably other interesting pathological cases not covered, however this kind of spectrum does include double-wells, multiple-wells and all the greatest hits of one-dimensional solvability like the harmonic well, infinite square well,  P\"{o}lsch-Teller potential, Morse potential, etc.

Eigenstates of $\hat{H}^1$ are denoted by kets containing the spectral index
\begin{equation}
\hat{H}^1\kt{n} = \epsilon_n \kt{n}.
\end{equation}
and the corresponding wave functions are
\begin{equation}
\phi_{n}(q)=\bk{q}{n}.
\end{equation}
No functional dependence of $\epsilon_n$ on $n$ is implied, although algebraic or transcendental expressions certainly exist for specific solvable potentials. For convenience, sometimes the one-particle eigenstates will be denoted by state labels $\kt{\alpha}$, $\kt{\beta}$, $\kt{\gamma}$, etc.\ with wave functions $\phi_\alpha(q)$ and (for symmetric traps) parities $\pi_\alpha$.

\subsection{Configuration Space Symmetries for One Particle}

The configuration space symmetry $\mathrm{C}_1$ is the group of all transformations of $\mathcal{Q}^1=\mathbb{R}$ realized by operators that commute with the one-particle Hamiltonian $\hat{H}^1$. For an asymmetric trap, no such operators exist and $\mathrm{C}_1 \sim \mathrm{Z}_1$ is the trivial group.

For a symmetric trap, there is a single point about which reflections are a symmetry and $\mathrm{C}_1 \sim \mathrm{O}(1)$ is the parity group. For symmetric one-dimensional wells, the quantum number $n$ also determines the parity 
\begin{equation}
\hat{\Pi}\kt{n} = (-1)^n \kt{n}.
\end{equation}

Although not a trap (and outside the purview of this article), for a constant potential (e.g.\ no potential) the group $\mathrm{C}_1$ is the Euclidean group in one dimension $\mathrm{E}_1 \sim \mathrm{O}(1) \ltimes \mathrm{T}_q$, where $\mathrm{T}_q\sim\mathbb{R}^1$ is the group of spatial translations in $\mathcal{Q}^1$ and $\ltimes$ denotes the semidirect product. See \cite{Kalozoumis2014} for a discussion of symmetries and partial symmetries of lattice-like multi-well potentials.

\subsection{Kinematic Symmetries for One Particle}

The one-particle kinematic symmetry group $\mathrm{K}_1$ is the group of \emph{all} unitary symmetry operators that commute with $\hat{H}^1$, and therefore necessarily contains $\mathrm{C}_1$.

\emph{Asymmetric Traps}: For asymmetric traps,  the only symmetry is time translation so $\mathrm{K}_1\sim\mathrm{T}_t$. The irreps are one-dimensional, consistent with the non-degeneracy of $\sigma_1$ and are labeled by the energy $\epsilon_n$ or quantum number $n$. The spatial Hilbert space can be decomposed into irreps of $\mathrm{K}_1$:
\begin{equation}\label{K1decomp}
\KHS = \bigoplus_{n=0}^\infty \mathcal{K}^n.
\end{equation}
Each summand $\mathcal{K}^n$ is the one-dimensional irrep of time translation where time evolution is represented as $\underline{D}^n(t)= \exp(-i\epsilon_n t)$.

\emph{Symmetric Traps}: For symmetric traps, the kinematic group is $\mathrm{K}_1\sim\mathrm{O}(1)\times\mathrm{T}_t$. Irreps are still one-dimensional and labeled by $n$. The decomposition of $\KHS$ into $\mathrm{K}_1$ irreps is the same as (\ref{K1decomp}), except now parity $\pi_n=(-1)^n$ is also a good quantum number. Therefore the spatial Hilbert space $\KHS$ also has a decomposition into sectors of fixed parity $\KHS=\KHS^+ \oplus \KHS^-$ where 
\[
\KHS^+ = \bigoplus_{k=0}^\infty \mathcal{K}^{2k}\ \mbox{and}\ \KHS^- = \bigoplus_{k=0}^\infty \mathcal{K}^{2k+1}.
\]

\emph{Harmonic Traps}: For harmonic traps $\mathrm{K}_1 \sim \mathrm{U}(1)\times\mathrm{T}_t$. Here $\mathrm{U}(1)$ is the group of transformations that changes the phase of the ladder operators $\hat{a}$ and $\hat{a}^\dag$. Define a unitary representation of $\mathrm{U}(1)$ by operators $\hat{U}(\phi)$ for $\phi\in[0,2\pi)$ such that  
\[
\hat{b} = \hat{U}(\phi) \hat{a} \hat{U}^\dag(\phi)  = \exp(i\phi) \hat{a}.
\]
This transformation leaves $\hat{H}^1=\hat{a}^\dag \hat{a} +1/2$ invariant and can be thought of as rotations in two-dimensional phase space. Again, there is no change to the decomposition of $\KHS$ into $\mathrm{K}_1$ irreps (\ref{K1decomp}).

Note that for all three kind of traps, $\mathrm{C}_1$ and $\mathrm{K}_1$ are abelian groups. For abelian groups, irreducible representations are one-dimensional, and this is consistent with the assumption of a non-degenerate, discrete one-particle spectrum $\sigma_1$.

\subsection{Including Spin}

Finally, if the single-particle Hamiltonian $\hat{H}^1$ is spin independent  and there are $J$ spin components, then there is also a factor group of $\mathrm{U}(J)$ to the kinematic symmetry. The total Hilbert space for the one particle system is the tensor product of the spatial  Hilbert space and the spin Hilbert space.
\begin{equation}
\HS = \KHS \otimes \SHS \sim \mathrm{L}^2(\mathbb{R}) \otimes \mathbb{C}^J.
\end{equation}
Any unitary operator that acts only on $\SHS \sim \mathbb{C}^J$ certainly commutes with $\hat{H}^1$. Further, if the internal components really are spin components of a particle with spin $s$, then $J= 2s +1$ and the spin operators $\hat{S}^2$ and $\hat{S}_z$ form a complete set of commuting operators for $\SHS$ that commute with the Hamiltonian.

\section{Two-Particle Systems}

The purpose of this section is to classify the types of symmetries found for two trapped particles in the case of no interaction, a general two-body interaction, and the contact interaction. In some sense, symmetry analysis does not provide anything remarkable or new for two particles. However, it provides a training ground for intuition about symmetries in a familiar setting and it is useful for contrast with more complex scenarios. Also, techniques and notation are introduced here that are be extended to the three particle case in the next section, and then to the $N$-particle case in the sequel article.

One of the most important ideas of this section that the degeneracy of the two-particle spectrum $\sigma_2$ can be explained by looking at the kinematic symmetry group $\mathrm{K}_2$. The dimensions of $\mathrm{K}_2$-irreps should be the same as the degeneracies in $\sigma_2$. If not, that could signal the presence of an emergent two-particle symmetry, i.e.\ a symmetry that cannot be generated from one-particle symmetries and particle permutations.

\subsection{Two Non-Interacting Particles}

\begin{figure}
\centering
\includegraphics[width=.5\linewidth]{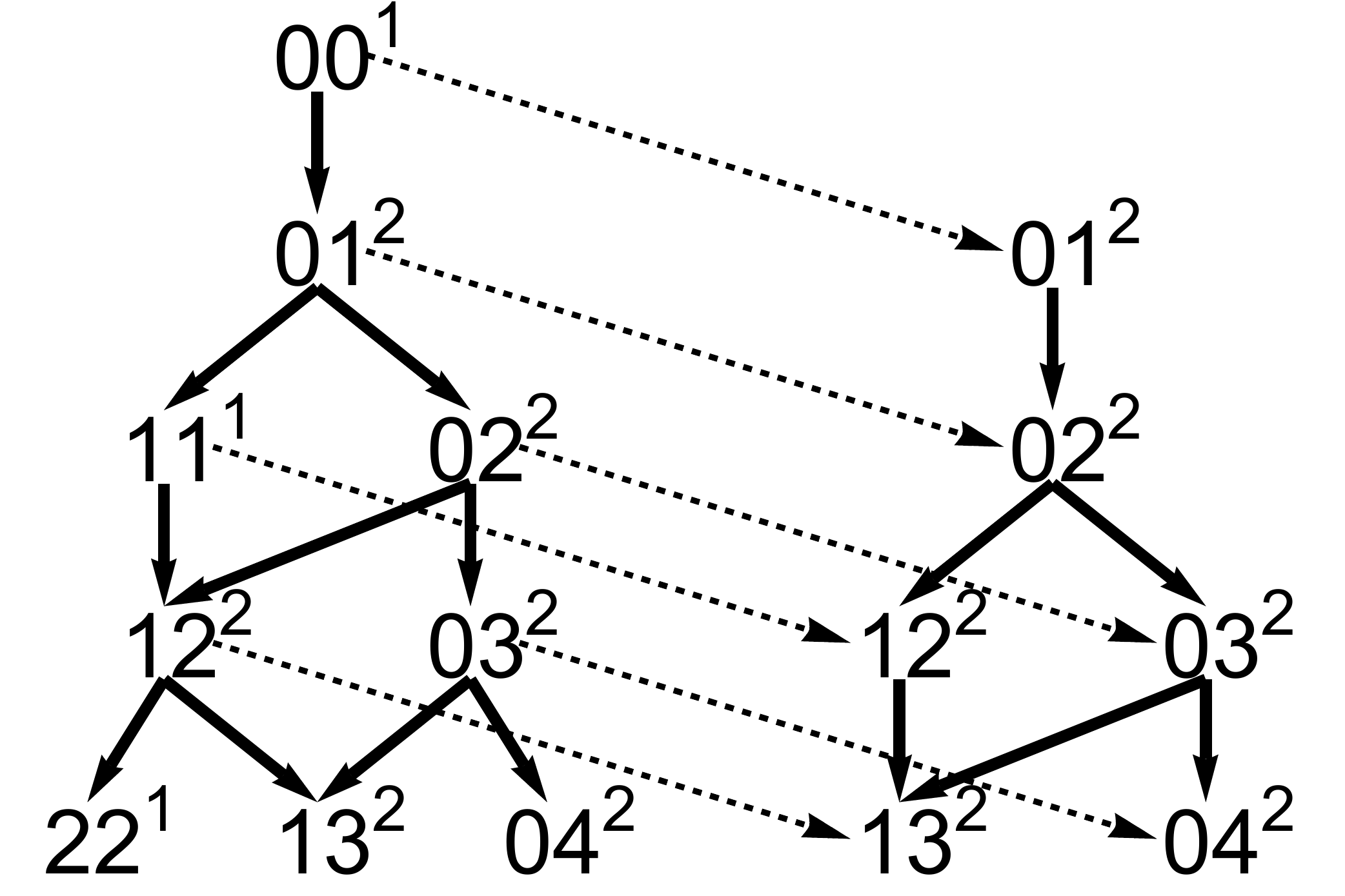}
\caption{On the left, each pair of one-particle quantum numbers numbers is a composition $\pp{\nu}$ corresponding to an energy $E_{\pp{\nu}}$ in the two-particle non-interacting spectrum $\sigma_2^0$. The superscript denotes that the compositions like $\pp{\alpha^2}$ with shape $[\nu]=[2]$ are non-degenerate and compositions like $\pp{\alpha\beta}$ with shape $[\nu]=[1^2]$ are two-fold degenerate. There is a totally symmetric spatial state in both shapes of compositions, but only compositions like $\pp{\alpha\beta}$ contain totally antisymmetric states. Note that there is a one-to-one map (depicted with dotted arrows) from all compositions levels in $\sigma_2^0$ (on left) to all mixed compositions $\pp{\alpha\beta}$ (on right), and the map preserves the partial ordering. This is one way to depict the famous boson-fermion mapping for two strongly-interacting particles in one dimension.}
\label{fig:partialorder2}
\end{figure}

Consider the total non-interacting Hamiltonian constructed from the sum of one-particle Hamiltonians
\begin{equation}\label{H02}
\hat{H}^2_0 = \hat{H}^1_1 + \hat{H}^1_2 = \frac{1}{2m}\left( \hat{P}_1^2 + \hat{P}_2^2 \right) + V^1(\hat{Q}_1) + V^1(\hat{Q}_2).
\end{equation}
The two-particle, non-interacting spectrum, denoted $\sigma_2^0$, remains discrete and countably-infinite, but unlike the one particle spectrum $\sigma_1$ it is necessarily degenerate. Every energy $E_{\pp{\alpha\beta}} \in \sigma_2^0$ is associated to (at least) one composition $\pp{\alpha\beta}$ of two energies $\epsilon_\alpha,\epsilon_\beta\in\sigma_1$. Unless the specific values of one-particle energies are known, only a partial ordering of $\sigma_2^0$ is possible. The lowest two energies in $\sigma_2^0$ are unambiguous: $E_{\pp{00}}=2\epsilon_0$ and $E_\pp{01}=\epsilon_0 + \epsilon_1$. However, the comparison of $E_{\pp{11}}$ and $E_\pp{02}$ is not possible without specific knowledge of the values for $\epsilon_0$, $\epsilon_1$ and $\epsilon_2$. For example, consider the potential $V^1(q)=|q|^z$. For $z=2$ the spectrum is harmonic with $E_\pp{11}=E_\pp{02}$, for $0<z<2$ the spectrum is softer than harmonic with $E_\pp{11} >E_\pp{02}$, and for $z>2$ the spectrum is harder than harmonic with $E_\pp{11} < E_\pp{02}$. See Fig.~\ref{fig:partialorder2} for a depiction of the partial ordering that can be put on $\sigma_2^0$ without specific knowledge of $\sigma_1$.

\subsubsection{Two Non-Interacting Particles: Configuration Space Symmetries}

\begin{figure}
\centering
\includegraphics[width=.7\linewidth]{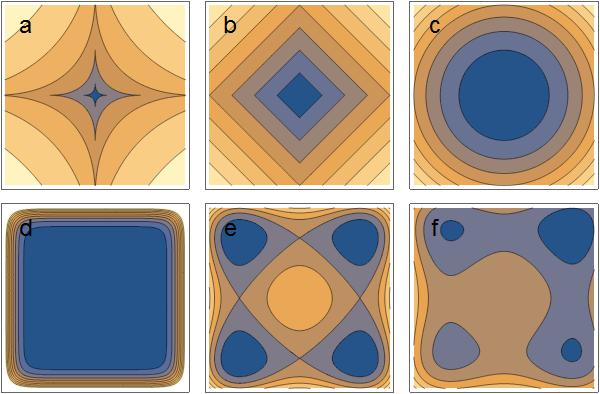}
\caption{These are the equipotentials for two particles in a one-dimensional trap with (a) $V(q)=|q|^{1/2}$ (cusped); (b) $V(q)=|q|$ (V-shaped); (c) $V(q)=q^2$ (harmonic); (d) $V(q)=q^{10}$ (approximately hard wall); (e) $V(q)=q^4 - 2 q^2$ (symmetric double well); and (f) $V(q)=q^4 - 1/3 q^3 - 2 q^2$ (asymmetric double well). The horizontal axis is the $q_1$ axis and the vertical axis is $q_2$. These figures all have particle exchange symmetry, which is realized in configuration space as a reflection across the line $q_1=q_2$. Antisymmetric spatial states must have a node on this line. Potentials (a)-(e) are also symmetric under spatial inversion of each particle individually, corresponding to horizontal and vertical reflections. For (a), (b), (d) and (e), the total symmetry group, combining parity and particle exchange is isomorphic to the non-abelian, two-dimensional point group $\mathrm{D}_4$. This order $8$ group is the symmetries of a square and includes the identity, rotation by $\pm \pi/2$, rotation by $\pi$, and reflection across three axes. The equipotentials of the harmonic well have the maximal point symmetry in two-dimensions: $\mathrm{O}(2)$, i.e.\ rotations by any angle and reflections across any axis.}
\label{fig:twopartequip}
\end{figure}

The configuration space for a system of two particles is $\mathcal{Q}^2=\mathbb{R}^2$. 
Fig.\ \ref{fig:twopartequip} depicts equipotentials for six traps, and without interactions this two-particle system is equivalent to one particle navigating these two-dimensional potentials. At a minimum, the configuration space symmetry group $\mathrm{C}_2^0$ for two identical, non-interacting particles  must contain as subgroups two copies of the one-particle symmetry group $\mathrm{C}_1$. The particle permutation group $\mathrm{P}_2\sim\mathrm{S}_2$ must also be a subgroup. This subsubsection establishes that the right way to combine these symmetries is
\begin{equation}
\mathrm{C}_2^0 \supseteq \mathrm{P}_2 \ltimes \mathrm{C}_1^{\times 2}.
\end{equation}

\emph{Asymmetric Trap}: The case with the absolutely minimum symmetry possible is the asymmetric trap. Then $\mathrm{C}_1\sim\mathrm{Z}_1$. and $\mathrm{C}^2_0 \sim \mathrm{S}_2$. Particle exchange $(12)$ acts on $\mathcal{Q}^2$ by
\[
\underline{O}(12)(q_1,q_2) = (q_2,q_1).
\]
This representation on $\mathcal{Q}^2$ is isomorphic two-dimensional point group denoted $\mathrm{D}_1$, the dihedral group generated by single reflection along the line $q_1=q_2$.

The spatial Hilbert space can be decomposed into subspaces corresponding to the irreps of $\mathrm{S}_2$: 
\begin{equation}\label{KHS:minimal2}
\KHS = \KHS^{[2]} \oplus \KHS^{[1^2]},
\end{equation}
where each of $\KHS^{[2]}$ is a tower of symmetric states and $\KHS^{[1^2]}$ is a tower of antisymmetric states.
The Hilbert space representation of particle exchange is the unitary operator $\hat{U}(12)$:
\begin{equation}
\hat{U}(12)\kt{\alpha\beta} = \kt{\beta\alpha}.
\end{equation}
Symmetric basis vectors $\kt{\alpha\,\alpha} \equiv \ykt{\tiny\young(\alpha\alpha)}$ are invariant under $\hat{U}(12)$ and are elements of the irrep tower $\KHS^{[2]}$.  However, the particle basis energy eigenstates $\kt{\alpha\beta}$ and $\kt{\beta\alpha}$  ($\alpha\neq\beta$)  do not belong to irrep towers. Instead, define the following simultaneous eigenvectors of $\hat{H}^2_0$ and  $\hat{U}(12)$:
\begin{eqnarray}\label{basis:K02}
\ykt{\tiny\young(\alpha\beta)} &\equiv& \frac{1}{\sqrt{2}} \left( \kt{\alpha\,\beta} +  \kt{\beta\,\alpha} \right) \in \KHS^{[2]}\nonumber\\
\ykt{\tiny\young(\alpha,\beta)} &\equiv& \frac{1}{\sqrt{2}}\left( \kt{\alpha\,\beta} -  \kt{\beta\,\alpha}\right) \in \KHS^{[1^2]}.
\end{eqnarray}

\emph{Symmetric Trap}: For a spatially symmetric trap, the one-particle configuration space group is parity $\mathrm{C}_1\sim\mathrm{O}(1)$. Without interactions, each particle can be independently spatially inverted. Denote each particle's inversion operator on $\mathcal{Q}^2$ by $\Pi_i$ such that $\Pi_1(q_1,q_2) = (-q_1,q_2)$ and 
$\Pi_2(q_1,q_2) = (q_1,-q_2)$. These symmetries are represented in the Hilbert space on the particle basis as
\begin{equation}
\hat{\Pi}_i \kt{n_1\,n_2} = (-1)^{n_i}\kt{n_1\,n_2}.
\end{equation}
Including these two operations, the symmetry group $\mathrm{C}_2^0$ for a symmetric trap has eight elements:
\begin{equation}
e, (12), \Pi_1, \Pi_2, (12)\Pi_1, (12)\Pi_2, \Pi_1\Pi_2, \mbox{and}\ (12)\Pi_1\Pi_2.
\end{equation}
This group is isomorphic to the point group of a square $\mathrm{D}_4$. The corresponding transformations on $\mathcal{Q}^2$ are
\begin{equation}
R(0), \Sigma(\pi/4), \Sigma(\pi/2), \Sigma(0), R(\pi/2), R(-\pi/2), R(\pi), \mbox{and}\ \Sigma(-\pi/4),
\end{equation}
where $R(\phi)$ is a rotation about the origin by $\phi$ and $\Sigma(\phi)$ is a reflection across the line making an angle $\phi$ with the $q_1$ axis. 
See Fig.~\ref{fig:twopartequip} and contrast the first five subfigures, which all have at least $\mathrm{D}_4$ symmetry\footnote{By $\mathrm{D}_4$, here I mean the two-dimensional point group, i.e.\ the dihedral group with four reflection axes that is the symmetry group of a square. Coxeter notation for this pure reflection group is $\mathrm{BC}_2$ or $[4]$. The same symbol $\mathrm{D}_4$ is also Sch\"{o}nflies notation for the three-dimensional point group with Coxeter notation $[4,2]^+$. These two groups  are isomorphic, but have different geometrical realizations. In the three-dimensional sense, the group $\mathrm{D}_4$  is an order eight group consisting of only rotations and no reflections. It can be visualized as the symmetries of a square parallelepiped with sides two-color checkered by an even number of checks. The Sch\"{o}nflies notation for the three-dimensional version of the reflection group $\mathrm{D}_4$ is $\mathrm{C}_{4v}$ and it is the symmetry of a square parallelepiped with the two square ends painted different colors.}, while the last subfigure only has $\mathrm{D}_1$ symmetry.

Note that the group $\mathrm{D}_4$ is not abelian, e.g.\ $(12)\Pi_1 = \Pi_2 (12)$. Therefore $\mathrm{D}_4$ is not isomorphic to the direct product $\mathrm{S}_2 \times \mathrm{O}(1) \times \mathrm{O}(1)$ which would be abelian. Instead it is isomorphic to 
\begin{equation}\label{wreath2}
\mathrm{D}_4 \sim  \mathrm{S}_2 \ltimes \left( \mathrm{O}(1) \times \mathrm{O}(1) \right)  \equiv \mathrm{S}_2 \ltimes \mathrm{O}(1)^{\times 2}.
\end{equation}
The notation $\ltimes$ stands for the semi-direct product and captures the fact that the particle exchange $(12)\in\mathrm{S}_2$ conjugates elements $\Pi_1$ and $\Pi_2$ and therefore acts as an automorphism of the normal, abelian  subgroup $\mathrm{O}(1)^{\times 2}$.
 
The group $\mathrm{D}_4$ has five irreducible representations~\cite{Hamermesh}, four unfaithful one-dimensional irreps denoted $A_1$, $A_2$, $B_1$ and $B_2$ and and one faithful two-dimensional irrep denoted $E$. The spatial Hilbert space can therefore be decomposed like
\begin{equation} \label{C20:sym}
\KHS = \KHS^{A_1} \oplus \KHS^{A_2} \oplus  \KHS^{B_1} \oplus  \KHS^{B_2} \oplus  \KHS^{E}.
\end{equation}
The first four irrep towers in (\ref{C20:sym}) contain one-dimensional irreps that have positive total parity $\Pi\equiv\Pi_1\Pi_2$ and the two-dimensional irrep $E$ has negative total parity.
As an example, the energy levels included in Fig.~\ref{fig:partialorder2} are categorized into irrep spaces of $\mathrm{C}_2^0$ in Table~\ref{tab:2nonintirreps}.

\begin{table}[t]
\caption{This table categorizes the energy levels in the two-particle, non-interacting spectrum $\sigma_2^0$ for a general symmetric trap into irreps of the non-interacting configuration space symmetry group $\mathrm{C}_2^0 \sim \mathrm{S}_2\ltimes\mathrm{O}(1)^{\times 2}$, into equivalence classes of irreps of the kinematic symmetry group $\mathrm{K}_2^0 \sim \mathrm{S}_2\ltimes( \mathrm{O}(1)\times \mathrm{T}_t  )^{\times 2}$, and into equivalence classes of irreps of the interacting symmetry group $\mathrm{K}_2 \sim \mathrm{S}_2\times\mathrm{O}(1)\times \mathrm{T}_t$. Non-interacting energy levels are labeled by their one-particle compositions. See text for notation for irreps, and see Fig.~\ref{fig:partialorder2} for the partial ordering of these energy levels if the specific values of the one-particle spectrum are unknown. Note that because $\mathrm{K}_2^0 \supseteq \mathrm{C}_2^0$ and $\mathrm{K}_2^0 \supseteq \mathrm{K}_2$, generally the $\mathrm{K}_2^0$ irreps are reducible with respect to $\mathrm{C}_2^0$ and $\mathrm{K}_2$.}
\centering
\label{tab:2nonintirreps}
\begin{tabular}{|c|c|c|c|c|}
\hline
Composition & Degeneracy  & $\mathrm{C}_2^0$ irreps &  $\mathrm{K}_2^0$ classes & $\mathrm{K}_2$ classes\\
\hline
$\pp{0^2}$ & 1  &  $A_1$ & $\pp{+^2}$ &  $[2]^+$\\
$\pp{01}$ & 2  & $E$ & $\pp{+-}$ & $[2]^-\oplus[1^2]^-$\\
$\pp{1^2}$ & 1 & $A_2$ & $\pp{-^2}$  &  $[2]^+$\\
$\pp{02}$ & 2  & $A_1 \oplus B_1$ & $\pp{+_1+_2}$ & $[2]^+\oplus[1^2]^+$\\
$\pp{12}$ & 2 & $E$ & $\pp{+-}$ & $[2]^-\oplus[1^2]^-$  \\
$\pp{03}$ & 2  & $E$ & $\pp{+-}$& $[2]^-\oplus[1^2]^-$ \\
$\pp{2^2}$ & 1  & $A_1$ & $\pp{+^2}$ &  $[2]^+$\\
$\pp{13}$ & 2  & $A_2 \oplus B_2$ & $\pp{-_1-_2}$ & $[2]^+\oplus[1^2]^+$\\
$\pp{04}$ & 2 & $A_1 \oplus B_1$ & $\pp{+_1+_2}$ & $[2]^+\oplus[1^2]^+$ \\
\hline
\end{tabular}

\end{table}

\noindent

\emph{Harmonic Trap}: The largest point symmetry possible in $\mathcal{Q}^2$ is $\mathrm{O}(2)$, all orthogonal transformations of the plane, i.e.\ reflections through and rotations about the origin. This is the configuration space symmetry for the harmonic potential. I call this an emergent symmetry because, unlike the cases of the asymmetric and symmetric traps, for a harmonic trap the non-interacting configuration space symmetry $\mathrm{C}_2^0$ cannot be generated by single particle symmetries and particle permutations.

The irreducible representations of $\mathrm{O}(2)$ are labeled by $m\in\mathbb{N}$ and they are one-dimensional for $m=0$ and two-dimensional for $m>0$. They correspond to the polar harmonics $\exp(i m \phi)$. One can think about this as `angular momentum' in configuration space and construct an observable $\hat{L}_{12} \sim i( \hat{a}_1 \hat{a}_2^\dag - \hat{a}_2 \hat{a}_1^\dag)$ out of ladder operators that commutes with $\hat{H}^2_0$. The degeneracy of the total energy total energy $E=\hbar\omega(X+1)$ is $d(E)= X+1$, so the dimensions of the irreps of $\mathrm{O}(2)$ are insufficient to explain the degeneracies of $\sigma_2^0$ for the harmonic trap. Explaining the total degeneracy requires considering the full kinematic symmetry of $\hat{H}^2_0$.

\subsubsection{Two Non-Interacting Particles: Kinematic Symmetries}

The previous section established that the minimal two-particle configuration space symmetry is given by
\begin{equation}\label{C20min}
\mathrm{C}_2^0 \supseteq \mathrm{P}_2 \ltimes \mathrm{C}_1^{\times 2}
\end{equation}
This expression relates the one-particle configuration space symmetry $\mathrm{C}_1$ to the two-particle non-interacting configuration space symmetry of $\hat{H}^2_0$ for both symmetric and asymmetric traps.

A similar situation holds for the kinematic symmetry of $\hat{H}^2_0$. For general symmetric and asymmetric traps, the minimal kinematic symmetry is 
\begin{equation}\label{K20min}
\mathrm{K}_2^0 \supseteq \mathrm{P}_2 \ltimes \mathrm{K}_1^{\times 2}.
\end{equation}
The one-particle kinematic symmetry $\mathrm{K}_1$ always includes time translation $\mathrm{T}_t$, so now there are two time translations, one generated by each particle's Hamiltonian $\hat{U}_j(t) = \exp(-i \hat{H}^1_j t)$. Since the particles are non-interacting, their clocks are not linked and their time lines are independent.  Total time evolution is also a symmetry of course, but for non-interacting particles it can be generated by single particle time evolutions $\hat{U}(t)=\hat{U}_1(t)\hat{U}_2(t)$. Note that the exchange operator does not commute with the one-particle time-translations; instead one finds $\hat{U}(12)\hat{U}_1(t) = \hat{U}_2(t)\hat{U}(12)$. This means $\mathrm{K}_2^0$ is not abelian and so its faithful irreps are not one-dimensional. 

\emph{Asymmetric Traps}: For asymmetric traps, $\mathrm{K}_1$ is just $\mathrm{T}_t$ and the minimal kinematic symmetry group is $\mathrm{K}_2^0 \sim \mathrm{S}_2 \ltimes \mathrm{T}_t^{\times 2}$.   The irreps of $\mathrm{K}_2^0$ are labeled by the state composition $\pp{\nu}$ and the irrep space is denoted $\KHS^\pp{\nu}$. 
The decomposition of the spatial Hilbert space into $\mathrm{K}_2^0$ irreps is the a direct sum over all compositions spaces:
\begin{equation}\label{K02irreps}
\KHS = \bigoplus_{\pp{\nu}\in\sigma_1 \times \sigma_1} \KHS^\pp{\nu}.
\end{equation}
Each $\KHS^\pp{\nu}$ is an energy eigenspace with energy $E_\pp{\nu}$. Unless there are emergent symmetries or accidental degeneracies, then each $E_\pp{\nu}$ is distinct.

The irreps of $\mathrm{K}_0^2$ fall into two equivalence classes. Compositions like $\pp{\nu}=\pp{\alpha^2}$ with shape $[\nu]=[2]$ have  one-dimensional irrep spaces $\KHS^\pp{\alpha^2}$ spanned by $\ykt{\tiny\young(\alpha\alpha)}$. Compositions $\pp{\nu}=\pp{\alpha\beta}$ with shape $[\nu]=[1^2]$ have two-dimensional representation spaces $\KHS^\pp{\alpha\beta}$ spanned by $\ykt{\tiny\young(\alpha\beta)}$ and $\ykt{\tiny\young(\alpha,\beta)}$. Note that since $\mathrm{K}_2^0 \supset \mathrm{C}_2^0$, irreps of $\mathrm{K}_2^0$ may be reducible with respect to $\mathrm{C}_2^0$, for example $\pp{\alpha\beta} \sim [2] \oplus [1^2]$.

\emph{Symmetric Traps}: The inclusion of parity symmetry does not change the irrep structure or change the decomposition (\ref{K02irreps}), but now there are five equivalence classes instead of two\footnote{These are not the same five irreps as $\mathrm{C}_0^2 \sim \mathrm{D}_4$, but the fact that the number of equivalence classes of $\mathrm{K}_N^0$ irreps is the same as the number of $\mathrm{C}_N^0$ irreps is valid for any $N$.}. These equivalence classes are: $\pp{+^2}$ and $\pp{-^2}$, compositions of two copies of the same state with even parity or odd parity; $\pp{+_1+_2}$ and $\pp{-_1-_2}$, compositions of two different states both with even parity or odd parity; and $\pp{+-}$, compositions of an even and odd state.
See Table~\ref{tab:2nonintirreps} for examples of how compositions spaces are sorted into  $\mathrm{K}_0^2$ irrep equivalence classes and reduced into  $\mathrm{C}_0^2$ irreps for low-energy compositions.

\emph{Harmonic Traps}: Two particles in a harmonic trap is the same as a two-dimensional isotropic harmonic oscillator, and so $\mathrm{K}_2^0 \sim \mathrm{U}(2)$. A representation of $\underline{u}\in\mathrm{U}(2)$ is defined by operators $\hat{U}(\underline{u})$ that act on the pair of one-particle annihilation operators $\hat{\bf a} = (\hat{a}_1 , \hat{a}_2)$ as
\begin{equation}
\hat{U}(\underline{u})\hat{\bf a}\hat{U}^\dag(\underline{u}) = \underline{u}\hat{\bf a}.
\end{equation}
These transformations leave the non-interacting Hamiltonian $\hat{H}^2_0 = \hbar\omega(\hat{\bf a}^\dag \hat{\bf a} + 1)$ invariant. The group $\mathrm{U}(2)$ is isomorphic to the set of all symplectic, orthogonal  transformations of four-dimensional phase space. The irreducible representations of $\mathrm{U}(2)$ are equivalent to the more familiar $\mathrm{SU}(2)$: they are finite-dimensional and labeled by an non-negative integer $X$. This quantum number is the same as the total excitation $X=n_1+n_2$ of a pair of oscillators\footnote{For $\mathrm{SU}(2)$ is standard to use $j=X/2$ as the label.}. The dimension of the irreducible representation is $d(\mathrm{U}(2);X)=X+1$. As a consequence, when $X>1$ there must be multiple compositions $\pp{\nu}$ with the same energy, not just the one- or two-fold degeneracy inherited from the one-particle symmetry via the subgroup $\mathrm{S}_2 \ltimes (\mathrm{U}(1) \times \mathrm{T}_t)^{\times 2}$. These degeneracies imply the existence of other operators besides those generated by $\mathrm{S}_N$ and $\mathrm{K}_1$ that commute with $\mathrm{H}^2_0$. Several inequivalent complete sets of commuting operators can be chosen and these correspond to the different coordinate systems in which the two-dimensional isotropic harmonic oscillator separates, i.e.\ cartesian, polar and elliptic~\cite{Boyer6}.

Another kind of emergent two-particle kinematic `symmetry' results from accidental degeneracies. The most famous of these are the Pythagorean degeneracies that occur for the infinite square well (see for example, \cite{Leyvraz}). These are not usually interpreted as symmetries because there is no corresponding (linear or non-linear) transformation on configuration space or phase space that induces a unitary representation on the whole Hilbert space\footnote{Of course, what one person calls an accidental degeneracy could be an undiscovered symmetry! It seems unlikely that after all this time that Pythagorean degeneracies will find a description in terms of configuration space or phase space transformations, but there may be other cases of accidental degeneracies waiting to be revealed as globally-defined emergent symmetries.}.

Technically one can construct operators which exploit the accidental degeneracy as a symmetry, but to do so requires knowledge of the spectrum. For example, the square well states\footnote{With the convention that the ground state has $n=0$, the energy of infinite square well is $\epsilon_n = \epsilon_0 (n + 1 )^2$.} with $(n_1,n_2)=(0,6)$, $(6,0)$ and $(4,4)$ span a three-dimensional subspace $\KHS^{50} = \KHS^\pp{06}\oplus \KHS^\pp{4^2}$ with energy $50 \epsilon_0$. One can define a family of operators isomorphic to $\mathrm{U}(3)$ that act unitarily on the three-dimensional energy eigenspace $\KHS^{50}$ and act as the  identity on the rest of the spatial Hilbert space $\KHS \ominus \KHS^{50}$. Those operators would realize the accidental degeneracy as a kinematic symmetry group.  However, the construction of such operators requires prior knowledge of the degeneracy instead of actually explaining how the degeneracy arises from the kinematic symmetry of the Hamiltonian and acts trivially on most of $\KHS$. It is therefore not as useful as a true emergent, global kinematic symmetry.

\subsection{Two Particles: General Two-Body Interactions}

Now add a two-body interaction to the Hamiltonian:
\begin{equation}
\hat{H}^2 = \hat{H}^2_0 + \hat{V}_{12}.
\end{equation}
Only Galilean-invariant two-particle potentials $\hat{V}_{12}$ are considered. The requirement of Galilean invariance can be summarized algebraically in terms of commutation relations:
\begin{eqnarray}\label{galsym}
 [\hat{\Pi},\hat{V}_{12}] =  [\hat{Q}_1,\hat{V}_{12}] &=& [\hat{Q}_2,\hat{V}_{12}] = [\hat{U}(12),\hat{V}_{12}] = 0\nonumber\\
 \mbox{and}\ [\hat{P}_1,\hat{V}_{12}] &=& -[\hat{P}_2,\hat{V}_{12}].
\end{eqnarray}
The second line of (\ref{galsym}) is equivalent to saying that the two-body interaction commutes with the center-of-mass motion. Combined with the first line of (\ref{galsym}), this implies that the interaction can be written as  $\hat{V}_{12} = V^2(\sqrt{2}|\hat{R}_1|)$, where $\hat{R}_1 = (\hat{Q}_1 - \hat{Q}_2)/\sqrt{2} $ is the normalized relative position coordinate.

\subsubsection{Two-Body Matrix Elements}

The condition $[\hat{U}(12),\hat{V}_{12}] = 0$ also implies the two-particle matrix elements of the interaction $\br{\alpha \beta} \hat{V}_{12} \kt{\gamma \delta}$ have the property
\begin{equation}
\br{\alpha \beta} \hat{V}_{12} \kt{\gamma \delta} = \br{\beta \alpha } \hat{V}_{12} \kt{\delta \gamma}\equiv v_{\pp{\alpha\gamma}\pp{\beta\delta}}.
\end{equation}
This notation for the matrix elements emphasizes that this amplitude is relevant for the state transitions $\alpha\leftrightarrow\gamma$ and $\beta\leftrightarrow\delta$. The one-particle basis can be chosen so that these matrix elements are all real. The group $\mathrm{P}_2$ is a symmetry for both $\hat{H}^2_0$ and $\hat{V}_{12}$, so it remains a symmetry of the total interacting Hamiltonian $\hat{H}^2$. Therefore there are only matrix elements between states carrying the same irreducible representation of $\mathrm{S}_2$, i.e.\
\begin{equation}\label{sr:V12}
\ybr{\tiny\young(\alpha\beta)} \hat{V}_{12} \ykt{\tiny\young(\gamma,\delta)} = \ybr{\tiny\young(\alpha\beta)} \hat{H}^2 \ykt{\tiny\young(\gamma,\delta)} = 0
\end{equation}
for any states $\alpha$, $\beta$, $\gamma$ and $\delta$.

First order perturbation theory gives the level splitting of the non-interacting states in the limit of weak interactions. In terms of the interaction matrix elements for the symmetrized states (\ref{basis:K02}), the level splittings are 
\begin{eqnarray}\label{sr:V12:sym}
\ybr{\tiny\young(\alpha\beta)} \hat{V}_{12} \ykt{\tiny\young(\alpha\beta)} = v_{\pp{\alpha^2}\pp{\beta^2}} + v_{\pp{\alpha\beta}^2}\nonumber\\
\ybr{\tiny\young(\alpha,\beta)} \hat{V}_{12} \ykt{\tiny\young(\alpha,\beta)} = v_{\pp{\alpha^2}\pp{\beta^2}} - v_{\pp{\alpha\beta}^2},
\end{eqnarray}
where for brevity I denote $v_{\pp{\alpha^2}\pp{\beta^2}} = v_{\pp{\alpha\alpha}\pp{\beta\beta}}$ and $v_{\pp{\alpha\beta}^2}=v_{\pp{\alpha\beta}\pp{\alpha\beta}}$. This implies the familiar result that for two-particle interactions the interference between the direct channel and the exchange channel generally shifts the symmetrized state  more than than the antisymmetric state.

\subsubsection{Symmetries of Two Interacting Particles}

The minimal non-interacting symmetry $\mathrm{P}_2\ltimes\mathrm{K}_1^{\times 2}$  is partially broken by $\hat{V}_{12}$. The particle exchange symmetry $\mathrm{P}_2$ is preserved. The diagonal subgroup of $\mathrm{K}_1^{\times 2}$, i.e.\ elements like $\hat{\Pi}_1\hat{\Pi}_2$ and $\hat{U}_1(t)\hat{U}_1(t)$, still commutes with $\hat{H}^2_0$. Therefore the kinematic symmetry of the interacting two particle system $\mathrm{K}_2$ always contains a subgroup isomorphic to the one-particle kinematic symmetry $\mathrm{K}_1$, so:
\begin{equation}\label{K2:subset}
\mathrm{K}_2 \supseteq  \mathrm{P}_2 \times \mathrm{K}_1.
\end{equation}

\emph{Asymmetric Trap}: The minimal total kinematic symmetry of the interacting system in an asymmetric trap is $\mathrm{K}_2 \sim \mathrm{S}_2 \times \mathrm{T}_t$ and the configuration space symmetry is just $\mathrm{C}_2 \sim \mathrm{S}_2$. Both of these groups are abelian with only one-dimensional irreps, and so in this minimal case, the spatial Hilbert space decomposes into irrep towers $\KHS = \KHS^{[2]} \oplus \KHS^{[1^2]}$. Each  energy in $\sigma_2$ is non-degenerate and associated to either the $\mathrm{S}_2$ irrep $[2]$ or $[1^2]$, unless the interacting system has emergent symmetries.

\emph{Symmetric Trap}: If the trap respects parity then $\mathrm{K}_1 \sim  \mathrm{O}(1) \times \mathrm{T}_t$. This group has the same irreps as the asymmetric case, but double the number of irrep equivalence classes, so $\KHS$ is decomposable into four towers
\[
\KHS = \KHS^{[2]+} \oplus \KHS^{[2]-} \oplus \KHS^{[1^2]+} \oplus \KHS^{[1^2]-}.
\]
The total parity operator $\hat{\Pi}$ commutes with $\hat{V}_{12}$ and parity remains a good quantum number even when interactions are turned on. Therefore, in addition to selection rules against transitions between states with different exchange symmetries (\ref{sr:V12}), matrix elements of $\hat{V}_{12}$ between two-particle states with different parity must be also be zero. This reduces the number of matrix elements required for exact diagonalization in a truncated Hilbert. See Table~\ref{tab:2nonintirreps} for the reduction of $\mathrm{K}_2^0$ irreps into $\mathrm{K}_2$ irreps for the case of symmetric traps.

Note however that the one-particle parities $\hat{\Pi}_i$ do not commute with $\hat{V}_{12}$. For symmetric traps the configuration space symmetry is reduced from $\mathrm{C}^0_2\sim\mathrm{D}_4$ (with order eight and five irreps) to only  $\mathrm{C}_2\sim\mathrm{D}_2$ (with order four and four one-dimensional irreps). For two particles, the permutation operator $\hat{U}(12)$ can also be interpreted as relative parity: reflection across the line $q_1= q_2$ reverses relative position $\hat{U}(12) \hat{R}_1 = - \hat{R}_1\hat{U}(12)$. The  operator $\hat{U}(12)\hat{\Pi}$ is a reflection across the line $q_1 = - q_2$ that reverses the normalized center-of-mass position $\hat{R}= ( \hat{Q}_1 + \hat{Q}_2)/\sqrt{2}$ and leaves the relative position $\hat{R}_1$ invariant.

\emph{Harmonic Trap}: For the harmonic trap $\mathrm{K}_2 \sim  \mathrm{S}_2 \times \mathrm{U}(1)\times \mathrm{T}_t$ and the extra $\mathrm{U}(1)$ symmetry provides an additional good quantum number: the center-of-mass excitation $n$. The spatial Hilbert space $\KHS$ is decomposable into an infinite number of equivalence classes, one for each value of $n$, and these each further separate into parity towers: 
\[
\KHS = \bigoplus_{n \in \mathbb{N}} \left( \KHS^{[2]n} \oplus \KHS^{[1^2]n} \right).
\]
The total parity of states in $\KHS^{[2]n}$ is $\pi=(-1)^n$ and states in $\KHS^{[1^2]n}$ have parity $\pi=-(-1)^n$.
The selection rules preclude non-zero matrix elements among states in different towers $\KHS^{[\mu]n}$, and this makes a significant reduction in effort for calculating higher-order terms in a perturbation series or for making exact diagonalization in truncated Hilbert spaces.

In summary, the minimal kinematic symmetry $\mathrm{K}_2$ of the two-particle interacting Hamiltonian $\hat{H}^2$ is $\mathrm{P}_2  \times  \mathrm{K}_1$. This group has only one-dimensional irreps, and so unless there is an accidental or emergent symmetries\footnote{Another example of emergent symmetries is the case of harmonic interactions $V^2(\sqrt{2}|\hat{R}_1|) \propto \hat{R}_1^2$ in a harmonic trap. This system has interacting kinematic symmetry $\mathrm{K}_2 \sim (\mathrm{U}(1) \times \mathrm{T}_t)^{\times 2}$ because both the center-of-mass and relative coordinate act like a one-dimensional harmonic oscillator.}, there are no degeneracies in the interacting spectrum $\sigma_2$. The symmetry $\mathrm{K}_2$ is enough to completely specify the qualitative features of level splitting for weak interactions. To calculate the specific energy shift requires the two-particle interaction matrix elements of the form $v_{\pp{\alpha^2}\pp{\beta^2}}$ and $v_{\pp{\alpha\beta}^2}$, but the splitting is universal. 

\subsubsection{Two-Body Contact Interactions}

Now, specify $\hat{V}_{12}$ to be the contact interaction, which in the position representation is 
\begin{equation}\label{contacti}
V^2(|q_1-q_2|) = g \delta(q_1 - q_2) = \frac{1}{\sqrt{2}} g \delta(r_1),
\end{equation}
where $r_1=(q_1-q_2)/\sqrt{2}$ is the normalized relative position coordinate. This potential satisfies the Galilean invariance requirements (\ref{galsym}) and therefore the kinematic symmetry group contains at least the minimal symmetry $\mathrm{P}_2 \times \mathrm{K}_1$. The goal of this section is to find results that are trap-independent using symmetry methods alone. Note that the case of the contact interaction is analytically solvable for two-bodies for any value of $g$ in one-dimensional harmonic trap~\cite{avakian_1987, busch_two_1998, jonsell_interaction_2002} or for infinite square well~\cite{Oelkers2006,Gaudin1971}. Finding the energy for general $g$ requires solving a transcendental equation, but the system is integrable for both of these traps.

For the contact interaction, the two-particle matrix elements $\br{\alpha \beta} \hat{V}_{12} \kt{\gamma \delta} = v_{\pp{\alpha\gamma}\pp{\beta\delta}}$ are invariant under permutations of the four states $\alpha$, $\beta$, $\gamma$ and $\delta$. This is shown by
by going to the position representation where\footnote{Remember that wave functions of the one-particle Hamiltonian can be chosen as real without loss of generality.}
\begin{eqnarray}
 \br{\alpha \beta} \hat{V}_{12} \kt{\gamma \delta} &=& \int dq_1 dq_2 \psi_\alpha^*(q_1) \psi^*_\beta(q_2) g\delta(q_1 - q_2) \psi_\gamma(q_1) \psi_\delta(q_2)\nonumber\\
 &=& g \int dq \, \psi_\alpha(q) \psi_\beta(q) \psi_\gamma(q) \psi_\delta(q)\nonumber\\
 &\equiv& v_\pp{\alpha\beta\gamma\delta}.
\end{eqnarray} 
This `state permutation symmetry' of the contact interaction matrix elements means that in addition to zero matrix elements between states in different irreducible representation spaces of $\mathrm{S}_2$ as in (\ref{sr:V12}), the matrix element between totally antisymmetric states is also necessarily zero
\begin{equation}
\ybr{\tiny\young(\alpha,\beta)} \hat{V}_{12} \ykt{\tiny\young(\gamma,\delta)} =  0
\end{equation}
as one shows by inserting $v_{\pp{\alpha^2}\pp{\beta^2}} = v_{\pp{\alpha\beta}^2} = v_\pp{\alpha^2\beta^2}$ in (\ref{sr:V12:sym}). The consequence, as expected, is that  the fermionic states do not ``feel'' the contact interaction and remain stationary states of the Hamiltonian for all values of the interaction strength $g$, attractive $g<0$ or repulsive $g>0$.

\subsection{Two Particles: Unitary Limit of Contact Interactions}

In the unitary limit $g \rightarrow \infty$, the contact interaction is like a sword through configuration space, severing the two halves with a nodal line that no probability current can penetrate. Each sector in configuration space acts as a disjoint domain for wave functions. The particles are either in the specific left-to-right order $q_1 < q_2$ or in the order $q_2 < q_1$. The spectrum in the unitary limit $\sigma_2^\infty$ is therefore the same as the spectrum of totally-antisymmetric non-interacting states $\ykt{\tiny\young(\alpha,\beta)}$. There is a two-fold degenerate level $\sigma_2^\infty$ for every state pair $\alpha \neq \beta$ (see the right side of Fig.\ \ref{fig:partialorder2}). As in the non-interacting case, only a partial ordering of the spectrum $\sigma_2^\infty$ can be determined without specific knowledge of $\sigma_1$.

Define the `snippet' basis wave functions~\cite{Deuretzbacher2008, Fang2011} for each pair $\alpha \neq \beta$ and each order $q_i < q_j$ by
\begin{eqnarray}\label{bas:snip2}
\ybk{\bf q}{\tiny\young(\alpha,\beta);\{12\}} &=& \left\{ \begin{array}{cc} \sqrt{2} \bk{\bf q}{\Yvcentermath1\tiny\young(\alpha,\beta) \Yvcentermath0} = \phi_\alpha(q_1)\phi_\beta(q_2) - \phi_\alpha(q_2)\phi_\beta(q_1) & \mbox{for}\ q_1 < q_2\\
0 & \mbox{for}\ q_2 < q_1 \end{array} \right. \nonumber\\
\ybk{\bf q}{\tiny\young(\alpha,\beta);\{21\}} &=& \left\{ \begin{array}{cc} -\sqrt{2} \bk{\bf q}{\Yvcentermath1\tiny\young(\alpha,\beta) \Yvcentermath0} = \phi_\alpha(q_2)\phi_\beta(q_1) - \phi_\alpha(q_1)\phi_\beta(q_2) & \mbox{for}\ q_2 < q_1\\
0 & \mbox{for}\ q_1 < q_2 \end{array} \right.
\end{eqnarray}
The two vectors $\ykt{\tiny\young(\alpha,\beta);\{12\}}$ and $\ykt{\tiny\young(\alpha,\beta);\{21\}}$ form a basis for the energy eigenspaces of the unitary-limit Hamiltonian $\hat{H}^2_\infty$ with energy $\epsilon_\alpha + \epsilon_\beta$. From (\ref{bas:snip2}) they transform under particle exchange like
\begin{equation}
\hat{U}(12) \ykt{\tiny\young(\alpha,\beta);\{12\}} = \ykt{\tiny\young(\alpha,\beta);\{21\}}.
\end{equation}
The states that are symmetric and antisymmetric under particle exchange are
\begin{eqnarray}
\ykt{\tiny\young(\alpha,\beta);{\tiny\young(12)}} &=& \frac{1}{\sqrt{2}} \left( \ykt{\tiny\young(\alpha,\beta);\{12\}} +  \ykt{\tiny\young(\alpha,\beta);\{21\}} \right)\nonumber\\
\ykt{\tiny\young(\alpha,\beta);{\tiny\young(1,2)}} &=& \frac{1}{\sqrt{2}} \left( \ykt{\tiny\young(\alpha,\beta);\{12\}} - \ykt{\tiny\young(\alpha,\beta);\{21\}}\right).
\end{eqnarray}
The state $\ykt{\tiny\young(\alpha,\beta);{\tiny\young(1,2)}}$ is in fact just the original non-interacting, antisymmetric basis vector $\ykt{\tiny\young(\alpha,\beta)}$ and the state $\ykt{\tiny\young(\alpha,\beta);{\tiny\young(12)}}$ is its symmetrized version. Although these two states have the same position probability density $|\ybk{\bf q}{\tiny\young(\alpha,\beta);\tiny\young(12)}|^2=|\ybk{\bf q}{\tiny\young(\alpha,\beta);\tiny\young(1,2)}|^2$, they will have different momentum distributions because of the cusp in $\ybk{\bf q}{\tiny\young(\alpha,\beta);\tiny\young(12)}$.

When the trap is parity symmetric and the one-particle states $\kt{\alpha}$ and $\kt{\beta}$ have parities $\pi_\alpha$ and $\pi_\beta$, then one infers from (\ref{bas:snip2}) that 
\begin{eqnarray}
\hat{\Pi}\ykt{\tiny\young(\alpha,\beta);{\tiny\young(12)}} &=& -\pi_\alpha\pi_\beta \ykt{\tiny\young(\alpha,\beta);{\tiny\young(12)}} \nonumber\\
\hat{\Pi}\ykt{\tiny\young(\alpha,\beta);{\tiny\young(1,2)}} &=& \pi_\alpha\pi_\beta \ykt{\tiny\young(\alpha,\beta);{\tiny\young(1,2)}}.
\end{eqnarray}
In other words, at unitarity the symmetric state always has opposite parity to the antisymmetric state from which it is constructed.

\subsubsection{Two Particles: Near Unitary Limit}

What about the not-quite-unitary limit? Consider a weak perturbation $\hat{T}$ of $\hat{H}_2^\infty$ that mimics the effect of not quite having an infinite barrier. Such operator would allow a little tunneling between the two sectors of configuration space, and should decrease the energy of states like $\ykt{\tiny\young(\alpha,\beta);{\tiny\young(12)}}$ due to the less dramatic cusp at the nodal line $q_1=q_2$. It also must have zero matrix elements between antisymmetric states because those states do not feel the contact interaction. An operator $\hat{T}$ that satisfies those requirements has uniform matrix elements in the snippet basis: 
\begin{equation}
\ybr{\tiny\young(\alpha,\beta);\{12\}} \hat{T} \ykt{\tiny\young(\alpha,\beta);\{21\}} = \ybr{\tiny\young(\alpha,\beta);\{21\}} \hat{T} \ykt{\tiny\young(\alpha,\beta);\{12\}}  = \ybr{\tiny\young(\alpha,\beta);\{12\}} \hat{T} \ykt{\tiny\young(\alpha,\beta);\{12\}} = \ybr{\tiny\young(\alpha,\beta);\{21\}} \hat{T} \ykt{\tiny\young(\alpha,\beta);\{21\}} = -t. 
\end{equation}
For a specific trap with known energy eigenstates, the small positive constant $t$ can be calculated~\cite{Deuretzbacher2014} from the wave function $\ybk{\bf q}{\tiny\young(\alpha,\beta)}$ as
\begin{equation}
t = \frac{2}{g} \int_{-\infty}^{+\infty} dq_2  \left|\frac{\partial \ybk{\bf q}{\tiny\young(\alpha,\beta)}}{\partial q_1}\right|^2_{q_1=q_2}
\end{equation}
As expected, the eigenstates of $\hat{H}_2^\infty + \hat{T}$ are also the symmetrized $\ykt{\tiny\young(\alpha,\beta);{\tiny\young(12)}}$ and antisymmetrized $\ykt{\tiny\young(\alpha,\beta);{\tiny\young(1,2)}}$ states, now with eigenvalues $\epsilon_\alpha + \epsilon_\beta - 2t$ and $\epsilon_\alpha + \epsilon_\beta$, respectively.

As another application of this section, by combining the results for weak splitting from $\sigma_2^0$ and for not-quite-unitary splitting from $\sigma_2^\infty$, a one-to-one adiabatic mapping from non-interacting states to unitary states can be determined, the simplest case of the famous Fermi-Bose mapping~\cite{Girardeau1960}. The non-interacting symmetric state $\ykt{\tiny\young(\alpha\alpha)}$ is mapped to $\ykt{\tiny\young(\alpha,\beta);{\tiny\young(12)}}$ where $\beta=\alpha+1$ while the non-interacting symmetric state $\ykt{\tiny\young(\alpha\beta)}$ with $\alpha<\beta$ is mapped to $\ykt{\tiny\young(\alpha,\gamma);{\tiny\young(12)}}$ where $\gamma=\beta+1$. This amounts to adding one nodal line to each of the symmetric states at the location of the contact interaction. Of course the antisymmetric states $\ykt{\tiny\young(\alpha,\beta)}= \ykt{\tiny\young(\alpha,\beta);{\tiny\young(1,2)}}$ are unchanged under the adiabatic mapping because they already align with the nodal line at $q_1=q_2$.

\subsection{Spin and Symmetrization for Two Particles}

Before moving on from two particles, let us finally consider the incorporation of identical particle symmetrization and spin degrees of freedom and state some well-known results. When there are $J>1$ spin components accessible, then the total total Hilbert space $\HS = \KHS \otimes \SHS$ is the direct product of the spatial Hilbert space $\KHS\sim L^2(\mathbb{R}^2)$ and the spin Hilbert space $\SHS\sim \mathbb{C}^{J^2}$. The $\mathrm{P}_2$ symmetry implies that the total Hilbert space can be decomposed into symmetrized subspaces
\[
\HS = \HS^{[2]} \oplus \HS^{[1^2]}.
\]
If there are no spin degrees of freedom, then $\SHS$ is one-dimensional and the symmetric subspace of the spatial Hilbert space $\HS^{[2]} \sim \KHS^{[2]}$ is available for population by identical bosons and the antisymmetric subspace $\HS^{[1^2]} \sim \KHS^{[1^2]}$ by fermions. If there are spin degrees of freedom, then $\SHS$ can also be decomposed into symmetric and antisymmetric subspaces  $\SHS^{[2]}$ and $\SHS^{[1^2]}$. For example, for two particles with spin 1/2, the triplet states are in $\SHS^{[2]}$ and the singlet state is in $\SHS^{[1^2]}$. The spin Hilbert spaces and spatial Hilbert spaces are then combined as 
\begin{eqnarray}\label{hks2}
\HS^{[2]} &=& \left(\KHS^{[2]} \otimes \SHS^{[2]}\right) \oplus \left(\KHS^{[1^2]} \otimes \SHS^{[1^2]}\right)\ \mbox{and}\nonumber\\
\HS^{[1^2]} &=& \left(\KHS^{[2]} \otimes \SHS^{[1^2]}\right) \oplus \left(\KHS^{[1^2]} \otimes \SHS^{[2]}\right).
\end{eqnarray}
Note that the total spin operator $\hat{S}=(\hat{\bf S}_1 + \hat{\bf S}_2)^2$ and total spin component operator $\hat{S}_z=\hat{S}_{1z}+\hat{S}_{2z}$ commute with the permutation operator $\hat{U}(12)$, as well as with all one-particle and two-particle spatial observables. Therefore, total spin and spin component are good quantum numbers for the symmetrized states for any interaction as long as the trap is spin-independent.  Symmetrization induces correlations between spin states and energy, for example making the lowest energy state two spin-1/2 fermions only accessible to the singlet combination.

\section{Three Particles}

The kinematic symmetry of the Hamiltonian $\hat{H}^3$ (whether interacting or non-interacting)  includes particle permutation symmetry $\mathrm{P}_3 \sim \mathrm{S}_3$. Unlike $\mathrm{S}_2$, the group $\mathrm{S}_3$ is not abelian, and so now the irreducible representations of particle exchange symmetry are more complicated. One implication is that the spatial Hilbert space $\KHS$, the spin Hilbert space $\SHS$ and the total Hilbert space $\HS$ can each be broken into subspaces with the three types of symmetry that three particle states can have, e.g.\ for the spatial Hilbert space 
\begin{equation}\label{3sectors}
\KHS = \KHS^{[3]} \oplus \KHS^{[21]} \oplus \KHS^{[1^3]}.
\end{equation} 
Each of these subspaces $\KHS^{[\mu]}$ in (\ref{3sectors}) is a tower of irrep spaces of $\mathrm{S}_3$. If $\mathrm{S}_3$ is the only symmetry of $\hat{H}^3$, then each copy of each $\mathrm{S}_3$ irrep would have a distinct energy.

One complication of $\mathrm{S}_3$ compared to $\mathrm{S}_2$ is that the elements of $\mathrm{S}_3$ are not all their own inverses, and so now we have to be a little more careful about representations. I choose the convention that a particle permutation $p\in\mathrm{P}_3$ acts on the coordinates like
\begin{equation}\label{coordperm}
\underline{O}(p) (q_1,q_2,q_3) = (q_{p_1},q_{p_2},q_{p_3}),
\end{equation}
where $p$ is expressed in permutation notation $\{p_1 p_2 p_3\}$. Using $\Psi({\bf q})=\bk{\bf q}{\Psi}$, the induced representation on wave functions is
\[
\hat{U}(p)\Psi({\bf q}) = \Psi(\underline{O}(p^{-1}){\bf q}).
\]
This implies that  particle permutations are represented on the particle basis $\kt{n_1 n_2 n_3}$ similarly to eq.~(\ref{coordperm}):
\begin{equation}\label{partbasperm}
\hat{U}(p) \kt{n_1 n_2 n_3} = \kt{n_{p_1} n_{p_2} n_{p_3}}.
\end{equation}
This convention is opposite to the convention used in \cite{Chen}, which otherwise (especially chapters 3 and 4) provides an excellent reference for the methods used in this section and in the sequel.

Similar to the previous section, the following subsections treat the cases of non-interacting, interacting, and  contact interactions in the unitary limit. The usefulness of state permutation symmetry becomes more evident as the limits of particle permutation symmetry become more acute, and in the unitary limit of the contact interaction a new kinematic symmetry emerges called ordering permutation symmetry.

\subsection{Three Particles: Non-Interacting}

\begin{figure}
\centering
\includegraphics[width=.5\linewidth]{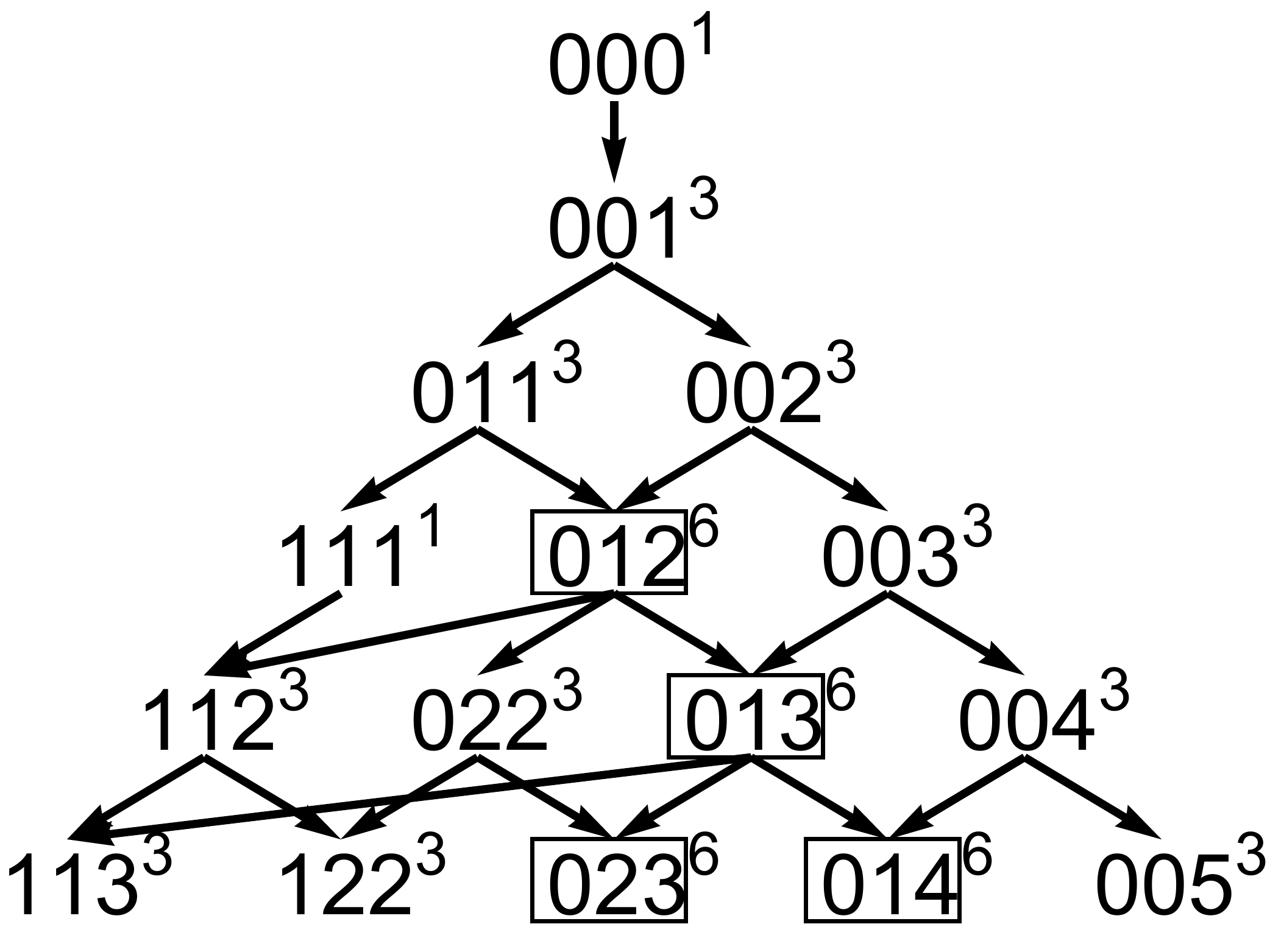}
\caption{Each sequence of three numbers determines a composition $\pp{\nu}$. The superscript denotes the number of sequences with that composition, or equivalently the dimension $d\pp{\nu}$ of $\KHS^\pp{\nu}$. Unless the specific values of $\epsilon_i\in\sigma_1$ are known, only the partial ordering of composition energies $E_\pp{\nu}$  given by the arrows is defined.  The boxed sequences have compositions with the shape $[1^3]$. Their composition subspaces $\KHS^\pp{\alpha\beta\gamma}$ carry the regular representation of $\mathrm{S}_3$ and have state permutation symmetry $\mathfrak{P}_\pp{\alpha\beta\gamma} \sim \mathrm{S}_3$. There is a one-component bosonic state in every composition space $\KHS^\pp{\nu}$, but there are only one-component fermionic states in spaces with the boxed compositions. Note that if the sequence $\langle 0,1,2\rangle$ is added element-wise to each of the original sequences (e.g.\ $\langle 0,0,2 \rangle + \langle 0,1,2 \rangle=\langle 0,1,4 \rangle$) the chart will have the same form, giving a one-to-one mapping from bosonic states to fermionic states, as in Fig.~\ref{fig:partialorder2}.}
\label{fig:partialorder3}
\end{figure}

The spectrum of three non-interacting particles $\sigma_3^0$ is constructed from all possible compositions of the single particle energies. As with two particles, the spectrum $\sigma_3^0$ can be only partially ordered without specific knowledge of $\sigma_1$ (see Fig.\ \ref{fig:partialorder3}). If there are no emergent or accidental symmetries, then there are three kinds of energy levels: singly-degenerate levels derived from compositions of identical states like $\pp{\alpha^3}$, three-fold degenerate levels from compositions of two different states like $\pp{\alpha^2\beta}$, and six-fold degenerate levels from compositions of three different states like $\pp{\alpha\beta\gamma}$. Since $\mathrm{S}_3$ has only  one- and two-dimensional irreps, it is clear that $\mathrm{S}_3$ symmetry alone cannot explain the degeneracies of $\sigma_3^0$.

The configuration space and kinematic symmetries of three non-interacting particles inherit the following subgroups by their tensor product construction:
\begin{eqnarray}
\mathrm{C}_3^0 &\supseteq& \mathrm{P}_3 \ltimes \mathrm{C}_1^{\times 3} \nonumber\\
\mathrm{K}_3^0 &\supseteq& \mathrm{P}_3 \ltimes  \mathrm{K}_1^{\times 3} .
\end{eqnarray}
Indeed the group $\mathrm{P}_3 \ltimes \mathrm{K}_1^{\times 3}$ has irreps that are one, three and six dimensional, as shown below. These irreps are labeled by the the three energies in the composition, i.e.\ the three characters of the time translation subgroup $\mathrm{T}_t^{\times 3}$. Therefore, the minimal kinematic symmetry is sufficient to explain the degeneracy of the non-interacting energy levels in $\sigma_3^0$ unless there are emergent symmetries or accidental degeneracies.

\subsubsection{Three Non-Interacting Particles: Configuration Space Symmetries}

\begin{figure}
\centering
\includegraphics[width=.7\linewidth]{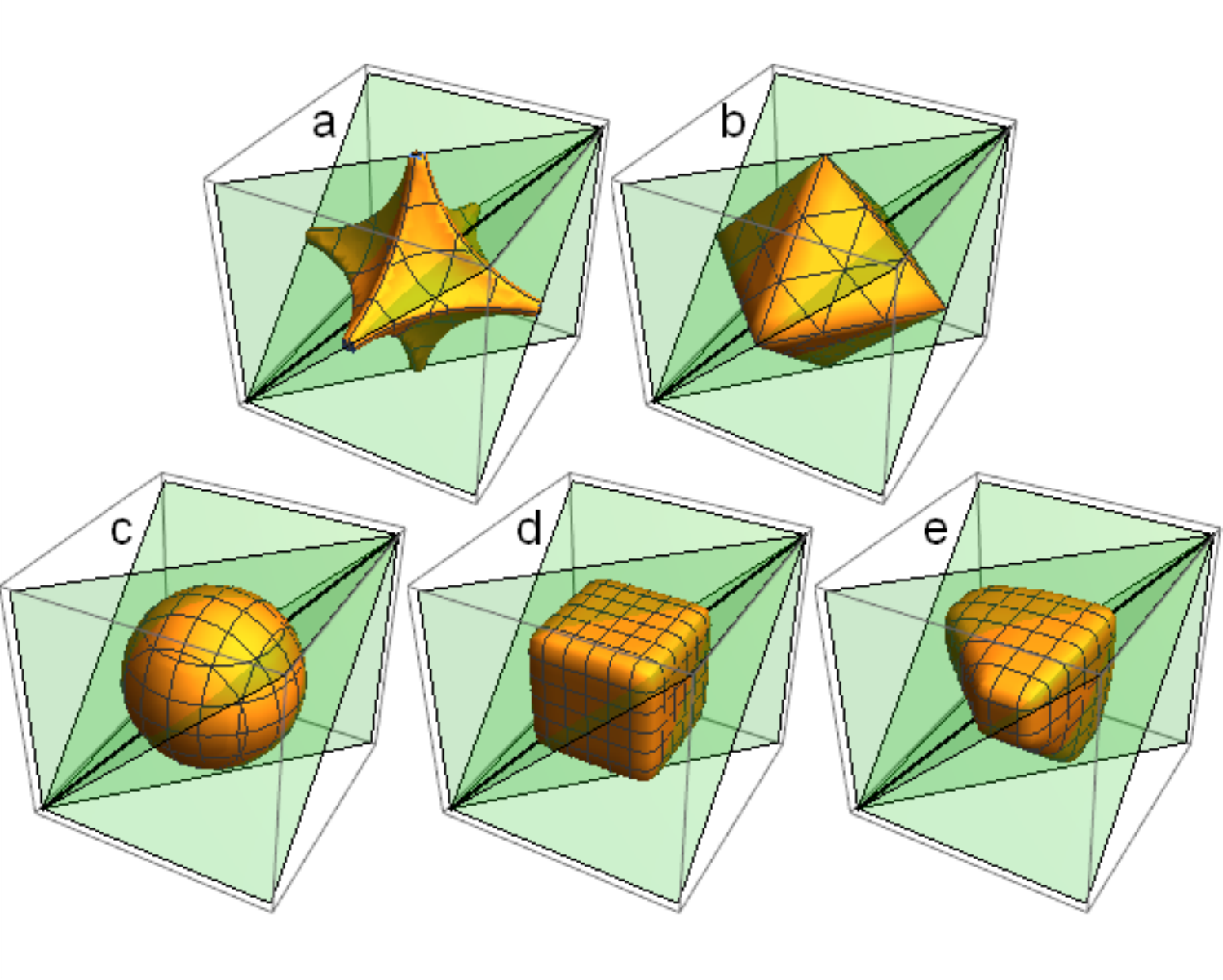}
\caption{These figures depict a single equipotential for three particles in a one-dimensional trap with (a) $V(q)=|q|^{1/2}$; (b) $V(q)=|q|$; (c) $V(q)=q^2$ (harmonic); (d) $V(q)=q^{10}$; (e) $V(q)=|q|$ for $q<0$ and $V(q)=q^{10}$ for $q>0$. The three transparent planes are the surfaces in configuration space where two particles coincide, i.e.\ $q_1=q_2$, $q_2=q_3$ and $q_1=q_3$. All subfigures have particle exchange symmetry $\mathrm{P}_3$: two-cycles like $(12)$ are reflections across the corresponding coincidence plane and three-cycles like $(123)$ are rotations by $2 \pi/3$ about the axis where all three planes intersect. Potentials (a)-(d) are also parity symmetric and have the full octahedral symmetry $\mathrm{O}_h$. Note that only (c) the harmonic potential (which has the maximal point symmetry $\mathrm{O}(3)$) is symmetric under relative parity inversion, corresponding to rotation of $\pi$ about the bold axis where the three coincidence planes intersect.}
\label{fig:equip3}
\end{figure}

\emph{Asymmetric Trap}: The minimal configuration space symmetry occurs when $\mathrm{C}_1\sim\mathrm{Z}_1$ and $\mathrm{C}_3^0 \sim \mathrm{S}_3$. The equivalent point group in three dimensions has Sch\"onflies notation $\mathrm{C}_{3v}$. This is the symmetry of a triangular prism with distinguishable ends. The permutations $p\in\mathrm{P}_3$ of three particles are realized in $\mathcal{Q}^3$ by an orthogonal matrix $\underline{O}(p)\in\mathrm{O}(3)$. Each of the three two-cycles $\underline{O}(ij)$ is a reflection across the plane defined by $q_i=q_j$ and the two three-cycles $\underline{O}(123)$ and $\underline{O}(132)$ are rotations by $\pm 2\pi/3$ about the line $q_1=q_2=q_3$.  See Fig.~\ref{fig:equip3} for some examples of equipotential surfaces for three non-interacting particles.

\emph{Symmetric Trap}: When the trap is parity symmetric, each particles' parity operator remains a symmetry of the system. Then $\mathrm{C}_1 \sim \mathrm{O}(1)$ and the configuration space symmetry is
\begin{equation}
\mathrm{C}_3^0 \sim \mathrm{S}_3 \ltimes \mathrm{O}(1)^{\times 3} \sim \mathrm{O}_h,
\end{equation}
where $\mathrm{O}_h$ is the full cubic symmetry in three-dimensions with order 48 and ten irreducible representations~\cite{Fernandez}, five with even parity and five with odd. See Table~\ref{tab:3nonintirreps} for a categorization of low-level non-interacting three particle states using the standard notation~\cite{Hamermesh} for the $\mathrm{O}_h$ irreps $A_{1g}$, $A_{2g}$, $E_g$, etc.

\begin{table}[t]
\caption{This table categorizes the energy levels in the three-particle, non-interacting spectrum $\sigma_3^0$ for a general symmetric trap into irreps of the non-interacting configuration space symmetry group $\mathrm{C}_3^0 \sim \mathrm{S}_3\ltimes\mathrm{O}(1)^{\times 3}\sim \mathrm{O}_h$, the kinematic symmetry group $\mathrm{K}_3^0 \sim \mathrm{S}_3\ltimes( \mathrm{O}(1)\times \mathrm{T}_t  )^{\times 3}$, and the interacting symmetry group $\mathrm{K}_3\sim\mathrm{S}_3\times\mathrm{O}(1)\times \mathrm{T}_t$. Non-interacting energy levels are labeled by their one-particle compositions. The notation for $\mathrm{O}_h$ irreps is standard, c.f.~\cite{Fernandez}, and the notation for the irreps of $\mathrm{K}_3^0$ is the composition with subscripts denoting each state's parity. See Fig.~\ref{fig:partialorder3} for the partial ordering of these energy levels if the specific values of the one-particle spectrum are unknown. One $\mathrm{C}_3^0$ irrep does not appear in this table; the lowest energy composition that carries a copy of the irrep $E_u$ is $\pp{1^23}\rightarrow A_{2u} \oplus E_u$. Two $\mathrm{K}_3^0$ irreps do not appear in this table; the lowest energy compositions that carry those irreps are $\pp{024}\rightarrow \pp{+_1+_2+_3}$ and $\pp{135}\rightarrow \pp{-_1-_2-_3}$.}
\centering
\label{tab:3nonintirreps}
\begin{tabular}{|c|c|c|c|c|}
\hline
Composition & Degeneracy  & $\mathrm{C}^3_0$ irreps &  $\mathrm{K}^3_0$ irreps & $\mathrm{K}^3$ irreps\\
\hline
$\pp{0^3}$ & 1  &  $A_{1g}$ & $\pp{+^3}$ &  $[3]^+$\\
$\pp{0^21}$ & 3  & $T_{1u}$ & $\pp{+^2-}$ & $[3]^-\oplus[21]^-$\\
$\pp{01^2}$ & 3 & $T_{2g}$ & $\pp{+-^2}$  &  $[3]^+\oplus [21]^+ $\\
$\pp{0^22}$ & 3  & $ A_{1g} \oplus E_g$ & $\pp{+_1^2+_2}$ & $[3]^+\oplus[21]^+$\\
$\pp{1^3}$ & 1 & $A_{2u}$ & $\pp{-^3}$ & $[1^3]^-$  \\
$\pp{012}$ & 6  & $T_{1u} \oplus T_{2u}$ & $\pp{+_1+_2-}$& $[3]^-\oplus 2 [21]^- \oplus [1^3]^-$ \\
$\pp{0^23}$ & 3  & $T_{1u}$ & $\pp{+^2-}$ & $[3]^-\oplus[21]^-$ \\
$\pp{1^22}$ & 3 & $T_{2g}$ & $\pp{+-^2}$  &  $[3]^+\oplus [21]^+ $\\
$\pp{02^2}$ & 3  & $ A_{1g} \oplus E_g$ & $\pp{+_1^2+_2}$ & $[3]^+\oplus[21]^+$\\
$\pp{013}$ & 6  & $T_{1g} \oplus T_{2g}$ & $\pp{+-_1-_2}$& $[3]^+\oplus 2 [21]^+ \oplus [1^3]^+$ \\
$\pp{0^24}$ & 3  & $ A_{1g} \oplus E_g$ & $\pp{+_1^2+_2}$ & $[3]^+\oplus[21]^+$\\
\hline
\end{tabular}

\end{table}

\emph{Harmonic Trap}: For a harmonic trap, the configuration space symmetry is maximal: $\mathrm{C}_3^0 \sim \mathrm{O}(3)$. As a result, another good basis of energy eigenstates is provided by the quantum numbers $\{X, \lambda, m\}$, where $X$ is total excitation, $\lambda\in\mathbb{N}$ labels the $\mathrm{O}(3)$ irrep and is like orbital angular momentum, and $m\in\{-\lambda,\ldots,\lambda\}$ labels an orthogonal basis within the irrep. Alternatively, the subgroup $\mathrm{O}(2)$ can be used to decompose the spatial Hilbert space into cylindrical harmonics and a center-of-mass quantum number. The cylindrical basis has proven particularly useful for exact diagonalization when interactions are included~\cite{Harshman2012, Garcia2014,Loft2014}.

\subsubsection{Three Non-Interacting Particles: Kinematic Symmetries}

As for the two-particle case, the irreps of the minimal kinematic symmetry $\mathrm{K}_3^0 \sim \mathrm{P}_3 \ltimes \mathrm{K}_1^{\times 3}$ are labeled by compositions $\pp{\nu}$. 
In the simplest case of an asymmetric trap, these irreps fall into three equivalence classes, depending on the shape of the composition. One basis for irreps spaces $\KHS^\pp{\nu}$ is the particle basis of all sequences in the composition, i.e.
\begin{eqnarray}\label{K30:hbasis}
\mathcal{K}^\pp{\alpha^3} &=& \mathrm{span}\{\kt{\alpha\alpha\alpha}\}\nonumber\\
\mathcal{K}^\pp{\alpha^2\beta} &=& \mathrm{span}\{\kt{\alpha\alpha\beta},\kt{\alpha\beta\alpha},\kt{\beta\alpha\alpha}\}\nonumber\\
\mathcal{K}^\pp{\alpha\beta\gamma} &=& \mathrm{span}\left\{\kt{\alpha\beta\gamma},\kt{\alpha\gamma\beta},\kt{\beta\alpha\gamma}, \kt{\beta\gamma\alpha},\kt{\gamma\alpha\beta},\kt{\gamma\beta\alpha}\right\}.
\end{eqnarray}
In the basis (\ref{K30:hbasis}), the subgroup $\mathrm{T}_t^{\times 3}$ has been diagonalized. A complete set of commuting observables for this basis are the generators $\{\hat{H}^1_1,\hat{H}^1_2,\hat{H}^1_3\}$. While this basis is natural, it is not optimized for the tasks of perturbation theory and exact diagonalization when interactions are incorporated, or for analyzing the unitary limit of contact interactions, or for symmetrization of identical fermions or bosons with or without spin.   For all these tasks, a basis optimized for the $\mathrm{S}_3$ subgroup works better. The rest of this subsubsection describes a particular choice for that basis and the corresponding complete set of commuting operators.

\emph{Asymmetric Trap}: Continuing with the asymmetric trap, the first step is to reduce the $\mathrm{K}_3^0$ irreps $\KHS^\pp{\nu}$ into $\mathrm{S}_3$ irreps labeled by $[\mu]$. For each of the three classes of $\mathrm{K}_3^0$ irrep spaces, the reduction looks like
\begin{eqnarray}\label{K3reduction}
\mathcal{K}^\pp{\alpha^3} &= & \mathcal{K}^{\tiny \young(\alpha\alpha\alpha)}\nonumber\\
\mathcal{K}^\pp{\alpha^2\beta} &= &  \mathcal{K}^{\tiny \young(\alpha\alpha\beta)} \oplus \mathcal{K}^{\tiny \young(\alpha\alpha,\beta)}\nonumber\\
\mathcal{K}^\pp{\alpha\beta\gamma} &= & \mathcal{K}^{\tiny \young(\alpha\beta\gamma)} \oplus \mathcal{K}^{\tiny \young(\alpha\beta,\gamma)} \oplus \mathcal{K}^{\tiny \young(\alpha\gamma,\beta)}  \oplus \mathcal{K}^{\tiny \young(\alpha,\beta,\gamma)}.
\end{eqnarray}
In each reduction, the irrep space $\KHS^\pp{\nu}$ is reduced into subspace $\KHS^W$ labeled by a semi-standard Weyl tableaux $W$. To make a Weyl tableaux, a Ferrers diagrams is filled with the state labels in the composition. These labels must stay the same or increase to the right and must increase to the bottom (assume $\alpha < \beta < \gamma$). There is only one way to fill a Weyl tableau out of the composition $\pp{\alpha^3}$, two ways for the composition $\pp{\alpha^2\beta}$, and four ways for $\pp{\alpha\beta\gamma}$. Each subspace $\KHS^W \subseteq \KHS^\pp{\nu}$ is isomorphic to the $\mathrm{S}_3$ irrep space $\mathcal{M}^{[W]}$ with the shape $[W]$ of the tableau $W$. Note that the dimensions of the irreps tally as they should in (\ref{K3reduction}), for example in the last line $d\pp{\alpha\beta\gamma} = d[3] + 2 d[21]  + d[1^3] = 6$. The proof of this reduction and the extension for $N$ particles is found in the sequel article. The key observation is that $\mathrm{K}_N^0$ irrep spaces $\mathcal{K}^\pp{\nu}$ are permutation modules~\cite{Sagan} of $\mathrm{S}_N$ characterized by the shape of the composition $[\nu]$. I now consider each equivalence class in turn.

First, irreps spaces like $\mathcal{K}^\pp{\alpha^3} = \KHS^{\tiny \young(\alpha\alpha\alpha)}$  carry the trivial, totally symmetric representation and have the single basis vector $\kt{\tiny \young(\alpha\alpha\alpha)}\equiv   \kt{\alpha\alpha\alpha}$. 

The second equivalence class of spaces $\KHS^\pp{\nu}$ are those like $\mathcal{K}^\pp{\alpha^2\beta}$. This representation (also called the defining representation) is reducible into a totally symmetric sector $\mathcal{K}^{\tiny \young(\alpha\alpha\beta)} \sim \mathcal{M}^{[3]}$ and a sector with mixed symmetry $\mathcal{K}^{\tiny \young(\alpha\alpha,\beta)} \sim \mathcal{M}^{[21]}$. Following \cite{Chen}, The three basis vectors of  $\mathcal{K}^\pp{\alpha^2\beta}$ can be chosen as
\begin{eqnarray}\label{basis:aab}
\ykt{\tiny \young(\alpha\alpha\beta)} &=& \frac{1}{\sqrt{3}} \left( \kt{\alpha\alpha\beta} + \kt{\alpha\beta\alpha} +  \kt{\beta\alpha\alpha}\right)  \nonumber\\
\ykt{\tiny \young(\alpha\alpha,\beta)  \,\young(12,3) }&=& \frac{1}{\sqrt{6}}  \left( 2 \kt{\alpha\alpha\beta} - \kt{\alpha\beta\alpha} -  \kt{\beta\alpha\alpha}\right) \nonumber\\
\ykt{\tiny \young(\alpha\alpha,\beta)  \,\young(13,2) }&=&  \frac{1}{\sqrt{2}} \left(  \kt{\alpha\beta\alpha} -  \kt{\beta\alpha\alpha}\right).
\end{eqnarray}
The second basis label for the $[21]$ irrep are the only two Young tableaux with the same shape as the Weyl tableau in the first basis label. When filling the Ferrers diagrams for a standard Young tableau, the particle numbers must increase to the right and to the bottom. The state $\ykt{\tiny \young(\alpha\alpha\beta)}$ does not need a Young tableau label because there is only one Young tableau $\tiny \young(123)$ with shape $[3]$, so it is understood.

These states (\ref{basis:aab}) are the simultaneous eigenvectors of $\hat{H}_3^0$ and two other operators:
\begin{equation}\label{cop3}
\hat{C}^\pp{123}_2 = \hat{U}(12) + \hat{U}(23) + \hat{U}(12)\ \mbox{and}\ \hat{C}^\pp{12}_2 = \hat{U}(12).
\end{equation}
The notation here seems a little chunky for three particles, but it can be generalized to $N$ particles and to other kinds of permutations, so I have adopted it. The operators $\hat{C}^\pp{\nu}_c$ are class operators over the all $c$-cycles of permutations of the composition $\pp{\nu}$. Specifically, $\hat{C}^\pp{123}_2$ is the sum of all two-cycles in $\mathrm{P}_3 \equiv \mathrm{P}_\pp{123}$ and $\hat{C}^\pp{12}_2$ is the only two-cycle in $\mathrm{P}_2 \equiv \mathrm{P}_\pp{12}$.
The basis vectors (\ref{basis:aab}) are eigenvectors of the operators in (\ref{cop3}):
\begin{eqnarray*}
\hat{C}^\pp{123}_2 \ykt{\tiny \young(\alpha\alpha\beta)} = 3 \ykt{\tiny \young(\alpha\alpha\beta)}, \ &&\ \hat{C}^\pp{12}_2 \ykt{\tiny \young(\alpha\alpha\beta)}=  \ykt{\tiny \young(\alpha\alpha\beta)},\nonumber\\
\hat{C}^\pp{123}_2 \ykt{\tiny \young(\alpha\alpha,\beta)  \,\young(12,3) } =0, \ &&\ \hat{C}^\pp{12}_2 \ykt{\tiny \young(\alpha\alpha,\beta)  \,\young(12,3) } =  \ykt{\tiny \young(\alpha\alpha,\beta)  \,\young(12,3) },\nonumber\\
\hat{C}^\pp{123}_2 \ykt{\tiny \young(\alpha\alpha,\beta)  \,\young(13,2) } =0, \ &&\ \hat{C}^\pp{12}_2 \ykt{\tiny \young(\alpha\alpha,\beta)  \,\young(13,2) } =  -\ykt{\tiny \young(\alpha\alpha,\beta)  \,\young(13,2) }.
\end{eqnarray*}
The two states $\ykt{\tiny \young(\alpha\alpha,\beta)\,\young(12,3) }$ and $\ykt{\tiny \young(\alpha\alpha,\beta)\,\young(13,2) }$ are therefore distinguished by whether particles $1$ and $2$ are symmetric or antisymmetric under exchange\footnote{This is equivalent to diagonalizing the $\mathrm{S}_3$ by the canonical subgroup chain $\mathrm{S}_3 \supset \mathrm{S}_2$ and up to a phase the basis (\ref{basis:aab}) is equivalent to the Yamanouchi basis~\cite{Chen,Hamermesh}.}. Therefore the set $\{\hat{H}_0^3,\hat{C}^\pp{123}_2,\hat{C}^\pp{12}_2\}$ are commuting operators that diagonalize composition spaces like $\mathcal{K}^\pp{\alpha^2\beta}$ with respect to the $\mathrm{S}_3$ subgroup of $\mathrm{K}_3^0$.

The third type of composition space $\mathcal{K}^\pp{\alpha\beta\gamma}$ is six-dimensional and carries the regular representation of $\mathrm{S}_3$, i.e.\ each representation appears as many times as its dimension. That means that the space $\mathcal{M}^{[21]}$ appears twice and the set of commuting operators $\{\hat{H}_3^0,\hat{C}^\pp{123}_2,\hat{C}^\pp{12}_2\}$ does not completely diagonalize $\mathcal{K}^\pp{\alpha\beta\gamma}$.

One way to solve this problem is to introduce state permutation group $\mathfrak{P}_\pp{\nu}$. The state permutation group $\mathfrak{P}_\pp{\alpha\beta\gamma}$ is isomorphic to $\mathrm{S}_3$ contains the six elements $e$, $(\alpha\beta)$, $(\beta\gamma)$, $(\alpha\gamma)$, $(\alpha\beta\gamma)$, and $(\alpha\gamma\beta)$. State permutations exchange state labels, not particle labels. Note the following comparisons that show they are distinct operations on $\mathcal{K}^\pp{\alpha\beta\gamma}$:
\begin{eqnarray}
 \hat{U}(12)\kt{\alpha\beta\gamma} = \kt{\beta\alpha\gamma} &=&  \hat{U}(\alpha\beta)\kt{\alpha\beta\gamma} = \kt{\beta\alpha\gamma}, \ \mbox{and}\ \nonumber\\
  \hat{U}(123)\kt{\alpha\beta\gamma} = \kt{\beta\gamma\alpha} &=&  \hat{U}(\alpha\beta\gamma)\kt{\alpha\beta\gamma} = \kt{\beta\gamma\alpha}, \ \mbox{but}\ \nonumber\\
\hat{U}(12)\kt{\beta\gamma\alpha} = \kt{\gamma\beta\alpha} & \neq&  \hat{U}(\alpha\beta)\kt{\beta\gamma\alpha} = \kt{\alpha\gamma\beta}, \ \mbox{and}\ \nonumber\\
\hat{U}(123)\kt{\beta\alpha\gamma} = \kt{\alpha\gamma\beta} &\neq &\ \hat{U}(\alpha\beta\gamma)\kt{\beta\alpha\gamma} = \kt{\gamma\beta\alpha}.
\end{eqnarray}
The elements $p\in\mathrm{P}_3$ and $\mathfrak{p}\in\mathfrak{P}_\pp{\alpha\beta\gamma}$ commute with each other on $\KHS^\pp{\alpha\beta\gamma}$, and there is a one-to-one map between the groups established by their action on the intrinsic state $\kt{\alpha\beta\gamma}$~\cite{Chen}.

State permutations are not a symmetries of $\mathrm{H}_3^0$. Exchanging two states changes in a particle basis vector usually changes the energy of the state. However, the state permutations $\mathfrak{P}_\pp{\nu}$ of a composition $\pp{\nu}$ are defined so that they leave the spaces $\mathcal{K}^\pp{\nu}$ invariant. For compositions like $\pp{\alpha^3}$ and $\pp{\alpha^2\beta}$, no state in those composition appears the same number of times and so state permutations are trivial $\mathfrak{P}_\pp{\alpha^3} \sim \mathfrak{P}_\pp{\alpha^2\beta} \sim \mathrm{Z}_1$. State permutations $\mathfrak{P}_\pp{\alpha\beta\gamma}$ distinguish the degenerate $[21]$ irreps of $\mathrm{S}_3$ in $\mathcal{K}^\pp{\alpha\beta\gamma}$ by introducing the operator
\begin{equation}\label{scop3}
\hat{C}^\pp{\alpha\beta}_2 = \hat{U}(\alpha\beta).
\end{equation}
Then, choosing the phase convention of \cite{Chen}, the following basis vectors are simultaneous eigenvectors of 
the set $\{\hat{H}_3^0,\hat{C}^\pp{123}_2,\hat{C}^\pp{12}_2, \hat{C}^\pp{\alpha\beta}_2\}$, operators which do not commute on all of $\KHS$, but do commute on $\mathcal{K}^\pp{\alpha\beta\gamma}$:
\begin{eqnarray}\label{basis:abg}
\ykt{\tiny \young(\alpha\beta\gamma)} &=& \frac{1}{\sqrt{6}} \left( \kt{\alpha\beta\gamma} + \kt{\beta\alpha\gamma} +  \kt{\gamma\beta\alpha}+  \kt{\alpha\gamma\beta}+  \kt{\gamma\alpha\beta}+  \kt{\beta\gamma\alpha}\right)\nonumber\\
\ykt{\tiny \young(\alpha\beta,\gamma)  \,\young(12,3)} &=& \frac{1}{\sqrt{12}} \left( 2\kt{\alpha\beta\gamma} + 2\kt{\beta\alpha\gamma} -  \kt{\gamma\beta\alpha}-  \kt{\alpha\gamma\beta} -  \kt{\gamma\alpha\beta} -  \kt{\beta\gamma\alpha}\right)\nonumber\\
\ykt{\tiny \young(\alpha\beta,\gamma)  \,\young(13,2)} &=& \frac{1}{2} \left( -  \kt{\gamma\beta\alpha} +  \kt{\alpha\gamma\beta} -  \kt{\gamma\alpha\beta} +  \kt{\beta\gamma\alpha}\right)\nonumber\\
\ykt{\tiny \young(\alpha\gamma,\beta) \, \young(12,3)} &=& \frac{1}{2} \left( -  \kt{\gamma\beta\alpha} +  \kt{\alpha\gamma\beta} +  \kt{\gamma\alpha\beta} -  \kt{\beta\gamma\alpha}\right)\nonumber\\ 
\ykt{\tiny \young(\alpha\gamma,\beta) \, \young(13,2)} &=& \frac{1}{\sqrt{12}} \left( 2 \kt{\alpha\beta\gamma} - 2 \kt{\beta\alpha\gamma} +  \kt{\gamma\beta\alpha}+  \kt{\alpha\gamma\beta} -  \kt{\gamma\alpha\beta} -  \kt{\beta\gamma\alpha}\right)\nonumber\\
\ykt{\tiny \young(\alpha,\beta,\gamma)} &=& \frac{1}{\sqrt{6}} \left( \kt{\alpha\beta\gamma} - \kt{\beta\alpha\gamma} -  \kt{\gamma\beta\alpha}-  \kt{\alpha\gamma\beta} +  \kt{\gamma\alpha\beta}+  \kt{\beta\gamma\alpha}\right).
\end{eqnarray}
As before, the the shape of the Weyl tableaux $W$ denote an irrep of $\mathrm{S}_3$ and the Young tableaux $Y\in [W]$ identify a basis for the irrep spaces. Particle permutations $p\in\mathrm{P}_3$ mix states with the same $W$ and different $Y$'s, and state permutations $\mathfrak{p} \in \mathfrak{P}_\pp{\alpha\beta\gamma}$ mix states with different $W$'s and the same $Y$. In Chapter 3 of \cite{Chen}, an explicit method for constructing a complete set of commuting operators that correspond to the $W$ and $Y$ is described for the general case of $N$ particles.

\emph{Symmetric Traps}: For symmetric traps, irreps of $\mathrm{K}_3^0$ are still labeled by compositions, but including parity information there are now ten equivalence classes. There are two equivalence classes $\pp{+^3}$ and $\pp{-^3}$ with the shape $[3]$; four classes $\pp{+_1^2 +_2}$, $\pp{+^2-}$, $\pp{+-^2}$ and $\pp{-_1^2-_2}$ with the shape $[21]$; and four classes $\pp{+_1+_2+_3}$, $\pp{+_1+_2-}$, $\pp{+-_1-_2}$ and $\pp{-_1-_2-_3}$ with the shape $[1^3]$. These reduce into $\mathrm{S}_3$ irreps the same way (\ref{K3reduction}) as the asymmetric trap. See Tab.~\ref{tab:3nonintirreps} for examples of low-energy compositions.

\emph{Harmonic Traps}: For the case of the harmonic trap, a state with total energy $E= \hbar\omega(X +3/2)$ has a degeneracy $d(\mathrm{U}(3);X)=(X+1)(X+2)/2$. These additional coincidences in $\sigma_3^0$ are explained by the emergent $\mathrm{U}(3)$ kinematic symmetry~\cite{Baker, Louck} and there are multiple inequivalent complete sets of commuting observables that separate the spatial Hilbert space.

\subsection{Incorporating Spin and Symmetrization}

For non-interacting spinless particles, the spectrum and degeneracy  is now effectively solved. Since  $\HS^{[\nu]}\sim\KHS^{[\nu]}$,  spinless distinguishable particles can populate every energy level in $\sigma_3^0$. Identical spinless bosons are restricted to the sector $\HS^{[3]}\sim\KHS^{[3]}$, where $\KHS^{[3]}$  is the tower composed of the single, spatially-symmetric  state that exists in every composition space $\mathcal{K}^\pp{\nu}$. Spinless fermions are restricted to the sector $\HS^{[1^3]}\sim\KHS^{[1^3]}$; spatially antisymmetric states exist only in composition spaces like $\KHS^\pp{\alpha\beta\gamma}$. There is a one-to-one mapping between the set of all composition spaces and the set of composition spaces with shape $[1^3]$; see Fig.\ \ref{fig:partialorder3}.

If there are internal components, there are two methods for symmetrization often used. One method fixes the spin components of specific particles, e.g.\ ``particle 1 and particle 2 are spin up and particle 3 is spin down.'' Then the spatial wave functions are symmetrized within particles with the same spin components. Examples of this approach include \cite{Hao2009, Guan2009, Guan2010, Harshman2012, Gharashi2013, Lindgren2014, Garcia2014, Levinsen2014, Loft2014, Gharashi2015}. The alternate method pursued here is the combined, simultaneous symmetrization of spin and spatial states, cf.\ \cite{Yurovsky2014, Yurovsky2015}. Following that approach, the spin Hilbert space $\SHS\sim \mathbb{C}^{J^N}$ can be decomposed into subspaces with definite symmetry
\[
\SHS = \SHS^{[3]} \oplus \SHS^{[21]} \oplus \SHS^{[1^3]}.
\]
Then the bosonic sector of $\HS$ is
\begin{equation}
\HS^{[3]} = (\KHS^{[3]} \otimes \SHS^{[3]} ) \oplus \left.( \KHS^{[21]}\otimes\SHS^{[21]})\right|_{[3]} \oplus ( \KHS^{[1^3]}\otimes \SHS^{[1^3]}),
\end{equation}
where $|_{[3]}$ means the one-dimensional, symmetric subspace of $\SHS^{[21]}\otimes \KHS^{[21]}$.
Bases for this subspace can be explicitly constructed using Clebsch-Gordan coefficients for $\mathrm{S}_3$, c.f.\ \cite{Chen, Hamermesh}. A similar expression for the fermionic sector is
\begin{equation}
\HS^{[1^3]} = ( \KHS^{[1^3]}\otimes \SHS^{[3]}) \oplus \left.( \KHS^{[21]}\otimes\SHS^{[21]})\right|_{[1^3]} \oplus ( \KHS^{[3]}\otimes\SHS^{[1^3]}).
\end{equation}
Note that $\SHS^{[1^3]}$ is non-empty only if $J\geq 3$; there must be at least three spin components in order for the spin state to carry the required antisymmetry to balance a totally symmetric spatial state.

Explicit state construction requires additional algebra, but counting degeneracies does not. As an example, consider three fermionic spin-$1/2$ particles with state labels $\uparrow$ and $\downarrow$ corresponding to the eigenvectors of the $z$-component of each particle's spin $\hat{S}_{i,z}$. The four states with total spin $s=3/2$ correspond to the totally symmetric spin vectors that span $\SHS^{[3]}$ that are labeled by the Weyl tableaux with shape $[3]$:
\[
\ykt{\tiny\young(\uparrow\uparrow\uparrow)}, \ykt{\tiny\young(\uparrow\downarrow\downarrow)}, \ykt{\tiny\young(\uparrow\uparrow\downarrow)}, \ykt{\tiny\young(\downarrow\downarrow\downarrow)}.
\]
In other words, for three spin-$1/2$ particles the space $\SHS^{[3]}$ carries one copy of the four dimensional $\mathrm{SU}(2)$ representation $\underline{D}^{s=3/2}$.
The four states with $s=1/2$ that span $\SHS^{[21]}$ can be chosen as simultaneous eigenvectors of $\hat{S}^2$, $\hat{S}_z$, and $\hat{U}(12)$:
\[
\ykt{\tiny\young(\uparrow\uparrow,\downarrow)\,\young(12,3)}, \ykt{\tiny\young(\uparrow\uparrow,\downarrow)\,\young(13,2)},\ykt{\tiny\young(\uparrow\downarrow,\downarrow)\,\young(12,3)},\ykt{\tiny\young(\uparrow\downarrow,\downarrow)\,\young(13,2)}.
\]
The first two have $\hat{S}_z$-eigenvalue $+1/2$, the second two  $-1/2$. The space $\SHS^{[21]}$ carries two copies of the $\mathrm{SU}(2)$ irrep $\underline{D}^{s=1/2}$. These two copies are distinguished by how they transform under $\hat{U}(12)$, as indicated by the Young tableau $\tiny \young(12,3)$ or $\tiny \young(13,2)$.

Combining these results with the spatial symmetries, for every composition space in the same class as $\KHS^\pp{\alpha\beta\gamma}$, there are four $s=3/2$ fermionic states from the product $ \KHS^{\tiny \young(\alpha,\beta,\gamma)} \otimes \SHS^{[3]}$ and four $s=1/2$ fermionic states from the reduction of the product $(\KHS^{\tiny \young(\alpha\beta,\gamma)}\oplus \KHS^{\tiny \young(\alpha\gamma,\beta)}) \otimes \SHS^{[21]} $. For every subspace like $\KHS^\pp{\alpha^2\beta}$ there are two $s=1/2$ fermionic states in the reduction of $\KHS^{\tiny \young(\alpha\alpha,\beta)} \otimes \SHS^{[21]}$. Three spin-$1/2$ fermions cannot populate energy levels like $\KHS^\pp{\alpha^3}$ because the spin Hilbert space cannot `carry' enough asymmetry to balance the symmetric state.

\subsection{Three Particles: General Interactions}

Now we add the pairwise interactions
\begin{equation}
\hat{H}^3 = \hat{H}^3_0 + \hat{V}^3,
\end{equation}
where
\begin{equation}
\hat{V}^3 = \hat{V}_{12} + \hat{V}_{23} + \hat{V}_{31}.
\end{equation}
In principle, there could also be an intrinsic three-body interaction that satisfies Galilean invariance and cluster decomposability, but here I only consider three-body interactions that result from pairwise interactions.

By construction, the operator $\hat{V}^3$ has $\mathrm{P}_3$ symmetry. It also inherits all the consequences of Galilean invariance from the two-particle interaction. Specifically, it commutes with the (normalized) center-of-mass position operator  and total momentum operator, as well as the total parity operator $\hat{\Pi}=\hat{\Pi}_1\hat{\Pi}_2\hat{\Pi}_3$. It also commutes with the relative parity operator $\hat{\Pi}_r$, which is an operator that cannot be generated from one-particle symmetries. For two particles, the relative parity operator acts identically to the permutation operator $\hat{U}(12)$, but for three particles relative parity is not an element of the particle permutation group $\mathrm{P}_3$ either. To see this, choose the particular set of normalized Jacobi coordinates
\begin{eqnarray}
\hat{R} &=& \frac{1}{\sqrt{3}}(\hat{Q}_1 + \hat{Q}_2 + \hat{Q}_3)\nonumber\\
\hat{R}_1 &=& \frac{1}{\sqrt{2}}(\hat{Q}_1 - \hat{Q}_2 )\nonumber\\
\hat{R}_2 &=& \frac{1}{\sqrt{6}}(\hat{Q}_1 + \hat{Q}_2 - 2  \hat{Q}_3).
\end{eqnarray}
Total parity inverts all three coordinates
\[
\hat{\Pi}\hat{R}=- \hat{R}\hat{\Pi}, \hat{\Pi}\hat{R}_1 =- \hat{R}_1\hat{\Pi}, \hat{\Pi}\hat{R}_2 = - \hat{R}_2\hat{\Pi} 
\]
whereas relative parity commutes with $\hat{R}$ and  inverts only the   relative positions
\[
\hat{\Pi}_r\hat{R} = \hat{R}\hat{\Pi}_r , \hat{\Pi}_r\hat{R}_1 =- \hat{R}_1\hat{\Pi}_r , \hat{\Pi}_r\hat{R}_2 =- \hat{R}_2\hat{\Pi}_r.
\]
In three-particle configuration space, $\Pi_r$ is realized as a rotation by $\pi$ around the line $q_1=q_2=q_3$.

\subsubsection{Three Interacting Particles: Configuration Space Symmetry}

Putting this together, the operator $\hat{V}^3$ has the configuration space symmetry group isomorphic to $\mathrm{S}_3 \times \mathrm{O}(1) \times (\mathrm{O}(1) \ltimes \mathrm{T}_R)$. The first factor is due to particle permutation symmetry, the second is total parity, and the third is translations and reflections along the center-of-mass axis. The trap will certainly break the translational symmetry of $\hat{V}^3$, and may also break other symmetries.

\emph{Asymmetric Trap}: For an general asymmetric trap, the configuration space symmetry remaining after two-body interactions are included is only $\mathrm{C}_3 \sim \mathrm{S}_3$; all parity symmetries are lost. This point group is the same as $\mathrm{C}_3^0$ for an asymmetric trap.

\emph{Symmetric Trap}: For a general symmetric trap, the total parity remains a good symmetry so $\mathrm{C}_3 \sim \mathrm{S}_3 \times \mathrm{O}(1)$. The Sch\"onflies notation for this three-dimensional point group with order twelve is $\mathrm{D}_{3d}$, Coxeter notation is $[[3]]$, and this is the symmetry of regular hexagonal prism with even-checkered sides. This is a subgroup of $\mathrm{C}_3^0 \sim \mathrm{O}_h$.

\emph{Harmonic Trap}: For a harmonic trap, relative parity provides another independent quantum number\footnote{Relative parity is also a good quantum number for uniform and linear traps because for any quadratic trap the center-of-mass and relative coordinates separate.}. The configuration space symmetry $\mathrm{C}_3 \sim \mathrm{S}_3 \times \mathrm{O}(1)  \times \mathrm{O}(1)$ is isomorphic to the three-dimensional point group $\mathrm{D}_{6h}$ and Coxeter notation is $[[3],2]$. This is the symmetries of a regular hexagonal prism and it is not a subgroup of $\mathrm{O}_h$.

\subsubsection{Three Interacting Particles: Kinematic Symmetry}

Unless there are emergent symmetries, the kinematic symmetry of $\hat{H}^3$ is $\mathrm{K}_3 \sim \mathrm{P}_3 \times \mathrm{K}_1$. The irreducible representations of this symmetry groups have the same dimensions as the irreducible representations of $\mathrm{S}_3$, independent of the particular one-particle symmetries $\mathrm{K}_1$. Every energy level $E\in\sigma_3$ is associated to an $\mathrm{S}_3$ irrep and  the spatial Hilbert space is decomposable into singly-degenerate levels for the totally symmetric irrep tower $\KHS^{[3]}$ and totally antisymmetric irrep tower $\KHS^{[1^3]}$ and doubly-degenerate energy levels for the irrep tower with mixed symmetry $\KHS^{[21]}$. Any other degeneracy pattern in $\sigma_3$ signals an emergent symmetry or accidental degeneracy.

One application of this decomposition is facilitating exact diagonalization in the non-interacting basis. Only basis vectors from the same $\mathrm{S}_3$ irreps with the same Young tableau will have non-zero interaction matrix elements:
\begin{equation}
\br{ W\, Y} \hat{V}^3 \kt{W' \, Y'} = \langle{W}||\hat{V}^3 || W'\rangle \delta_{YY'}.
\end{equation}
This allows the number of basis vectors needed to achieve a certain accuracy to be reduced.
For example, consider the 56 states that are depicted in Fig.~\ref{fig:partialorder3}. Of those, sixteen states are in $\KHS^{[3]}$ and only four states are in $\KHS^{[1^3]}$. The remaining 36 states are in $\KHS^{[21]}$, but since the interaction operator $\hat{V}^3$ acts like the identity within the irrep, only eighteen states are necessary for exact diagonalization.

If $\mathrm{K}_1$ includes parity symmetry, this provides an additional quantum number $\pi$.
Although parity does not change the degeneracy of the spectrum $\sigma_3$, it can be used to further decompose the spatial Hilbert space into two independent irrep towers for each $\mathrm{S}_3$ irrep, one for each parity:
\begin{equation}
\KHS= \KHS^{[3]+} \oplus \KHS^{[3]-} \oplus \KHS^{[21]+} \oplus \KHS^{[21]-} \oplus \KHS^{[1^3]+} \oplus \KHS^{[1^3]-}.
\end{equation}
Continuing the same example based on the states depicted in Fig.~\ref{fig:partialorder3}, the number of states needed to do exact diagonalization in each of these sectors is further reduced to 7, 9, 7, 11, 1, and 3, respectively.

For harmonic traps, a further reduction is possible. The factor of $\mathrm{U}(1)$ in $\mathrm{K}_1$ provides an additional quantum number conserved by the interactions: the center-of-mass excitation $n$. The irreps can be labeled by $[\mu]^\pm_n$. To calculate the spectrum $\sigma_3$ requires even fewer states because only the case $n=0$ needs to be calculated, e.g.\ there are only three basis states in Fig.~\ref{fig:partialorder3} that contribute to the ground state in irrep $[3]^+_0$. However, the double tableaux basis $\kt{W\, Y}$ is no longer the best basis for calculation. Instead, observables that exploit separability in cylindrical coordinates on $\mathcal{Q}^3$ oriented along the center-of-mass axis provide a more useful basis~\cite{Harshman2012,Harshman2014,Garcia2014,Loft2014}.

\subsubsection{Three Particles: Weak Interactions}

For weak interactions, the energy of the single state in each space like $\mathcal{K}^\pp{\alpha^3}$ shifts and the energy levels in the spaces like $\mathcal{K}^\pp{\alpha^2\beta}$ and $\mathcal{K}^\pp{\alpha\beta\gamma}$ split and shift. Unless the trap or interaction have additional symmetries (emergent or accidental), the degeneracy of the splitting is determined by the symmetry. Since $\mathrm{K}_3^0 \supset \mathrm{K}_3$, the irreps of $\mathrm{K}_3^0$ are generally reducible with respect to $\mathrm{K}_3$. In Table \ref{tab:3nonintirreps}, the reduction of composition spaces into $\mathrm{K}_3$ is given. This reduction has a similar form as the reduction of $\mathrm{K}_3^0$ irreps by $\mathrm{S}_3$ described by  (\ref{K3reduction}). For $\mathcal{K}^\pp{\alpha^3}$ and $\mathcal{K}^\pp{\alpha^2\beta}$ the reductions and splittings are identical, and no additional information about the nature of the two-body interaction or trap is required. However, for compositions like $\pp{\alpha\beta\gamma}$, the manner in which two copies of $\mathcal{M}^{[21]}$ split requires specific knowledge of the two-body matrix elements.

Let us make this explicit for each of the three types of composition spaces. In the particle basis, the matrix elements of $\hat{V}^3$ can be expressed in terms of the two-particle matrix elements:
\begin{equation}
\br{\alpha\beta\gamma}\hat{V}^3 \kt{\zeta\eta\theta} = v_{\pp{\alpha\zeta}\pp{\beta\eta}}\delta_{\gamma\theta} + v_{\pp{\beta\eta}\pp{\gamma\theta}}\delta_{\alpha\zeta} + v_{\pp{\alpha\zeta}\pp{\gamma\theta}}\delta_{\beta\eta}.
\end{equation}
Applying this, one-dimensional spaces like $\mathcal{K}^\pp{\alpha^3}$ experience an energy shift 
\begin{equation}
 \ybr{\tiny \young(\alpha\alpha\alpha)} \hat{V}^3 \ykt{\tiny \young(\alpha\alpha\alpha)} = 3 v_{\pp{\alpha\alpha}\pp{\alpha\alpha}} \equiv 3 v_{\pp{\alpha^2}^2}.
\end{equation}
The factor of three represents the fact that for this totally symmetric state the pairwise interactions of the three particles interfere constructively.

In three-dimensional spaces like $\mathcal{K}^\pp{\alpha^2\beta}$, the only non-zero matrix elements in the basis (\ref{basis:aab}) are 
\begin{eqnarray}
\ybr{\tiny \young(\alpha\alpha\beta)} \hat{V}^3 \ykt{\tiny \young(\alpha\alpha\beta)}  &=& v_{\pp{\alpha^2}^2} + 2 v_{\pp{\alpha^2}\pp{\beta^2}} + 2 v_{\pp{\alpha\beta}^2}\nonumber\\
\ybr{\tiny \young(\alpha\alpha,\beta) \, \young(12,3)} \hat{V}^3 \ykt{\tiny \young(\alpha\alpha,\beta)\, \young(12,3)} = \ybr{\tiny \young(\alpha\alpha,\beta) \, \young(13,2)} \hat{V}^3 \ykt{\tiny \young(\alpha\alpha,\beta)\, \young(13,2)}
&=& \rybr{\tiny \young(\alpha\alpha,\beta)} \hat{V}^3 \rykt{\tiny \young(\alpha\alpha,\beta)} \nonumber\\
&=& v_{\pp{\alpha^2}^2} + 2 v_{\pp{\alpha^2}\pp{\beta^2}} - v_{\pp{\alpha\beta}^2}.
\end{eqnarray}
The non-degenerate symmetric level always experiences a greater shift than the doubly degenerate mixed-symmetry level. Additionally, for contact interactions there is state permutation symmetry of the two-body matrix elements. Then we have $v_{\pp{\alpha^2}\pp{\beta^2}} = v_{\pp{\alpha\beta}^2} \equiv v_\pp{\alpha^2\beta^2}$ and these relations simplify further.

Can anything be considered `universal' for weak perturbations of composition spaces like $\mathcal{K}^\pp{\alpha^2\beta}$? Yes: the relation between the particle basis and the symmetrized perturbation eigenbasis, the larger energy shift of symmetric state compared to the partially symmetric state,  and the algebraic expression for the energy shift in terms of two-particle interaction matrix elements are all universal features of $\mathcal{K}^\pp{\alpha^2\beta}$ spaces. Although specific numerical values for the weak interaction energy shift depend on the specific two-body interactions, those properties do not.

For the $\mathcal{K}^\pp{\alpha\beta\gamma}$ composition spaces, one of these universal features is lost because such subspaces are not simply reducible by $\mathrm{S}_3$. The reduction of $\mathrm{K}_3^0$ irreps into $\mathrm{K}_3$ irreps is still sufficient to determine the level shifts in the totally symmetric and totally antisymmetric spaces. The matrix elements of $\hat{V}^3$ in terms of the two-body matrix elements are:
\begin{eqnarray}
\ybr{\tiny \young(\alpha\beta\gamma)} \hat{V}^3 \ykt{\tiny \young(\alpha\beta\gamma)}  &=& v_{\pp{\alpha^2}\pp{\beta^2}} +  v_{\pp{\alpha\beta}^2} + v_{\pp{\beta^2}\pp{\gamma^2}} +  v_{\pp{\beta\gamma}^2} + v_{\pp{\alpha^2}\pp{\gamma^2}} +  v_{\pp{\alpha\gamma}^2}\nonumber\\
\ybr{\tiny \young(\alpha,\beta,\gamma)} \hat{V}^3 \ykt{\tiny \young(\alpha,\beta,\gamma)} &=&  v_{\pp{\alpha^2}\pp{\beta^2}} -  v_{\pp{\alpha\beta}^2} + v_{\pp{\beta^2}\pp{\gamma^2}} -  v_{\pp{\beta\gamma}^2} + v_{\pp{\alpha^2}\pp{\gamma^2}} -  v_{\pp{\alpha\gamma}^2}.
\end{eqnarray}
These are the first-order energy shifts, and the first-order energy eigenstates remain $\ykt{\tiny \young(\alpha\beta\gamma)}$ and $\ykt{\tiny \young(\alpha,\beta,\gamma)}$  And as before, in the case of contact interaction the fermionic state will feel no energy shift at this (or any order) of perturbation theory. One way to think about this is destructive interference between the direct and exchange channels. Symmetry analysis gives another interpretation: because the interaction is symmetric under state permutation and the fermionic state is antisymmetric under state permutation, matrix elements between fermionic states are all identically zero.

However, for the two copies of the $\mathrm{S}_3$ irrep space $\mathcal{M}^{[21]}$ in the expansion of $\mathcal{K}^\pp{\alpha\beta\gamma}$, there are are matrix elements of $\hat{V}^3$ between states in $\KHS^{\tiny \young(\alpha\beta,\gamma)}$ and $\KHS^{\tiny \young(\alpha\gamma,\beta)}$ with the same Young tableau. The reduced matrix elements are 
\begin{eqnarray}\label{v3:abg}
\rybr{\tiny \young(\alpha\beta,\gamma) } \hat{V}^3 \rykt{\tiny \young(\alpha\beta,\gamma) } &=& v_{\pp{\alpha^2}\pp{\beta^2}} +  v_{\pp{\alpha\beta}^2}  + v_{\pp{\beta^2}\pp{\gamma^2}} - \frac{1}{2}  v_{\pp{\beta\gamma}^2} + v_{\pp{\alpha^2}\pp{\gamma^2}} - \frac{1}{2}  v_{\pp{\alpha\gamma}^2}   \nonumber\\
\rybr{\tiny \young(\alpha\gamma,\beta)} \hat{V}^3 \rykt{\tiny \young(\alpha\gamma,\beta) } &=& v_{\pp{\alpha^2}\pp{\beta^2}} -  v_{\pp{\alpha\beta}^2} + v_{\pp{\beta^2}\pp{\gamma^2}} + \frac{1}{2}  v_{\pp{\beta\gamma}^2} + v_{\pp{\alpha^2}\pp{\gamma^2}} + \frac{1}{2}  v_{\pp{\alpha\gamma}^2} \nonumber\\
\rybr{\tiny \young(\alpha\beta,\gamma) } \hat{V}^3 \rykt{\tiny \young(\alpha\gamma,\beta) } &=& \frac{\sqrt{3}}{2} (v_{\pp{\alpha\gamma}^2} -  v_{\pp{\beta\gamma}^2}).
\end{eqnarray}
Note that the state permutation symmetry of $\mathcal{K}^\pp{\alpha\beta\gamma}$ is broken by the interaction. The state permutation class operator $\hat{C}^\pp{\alpha\beta}_2 = \hat{U}(\alpha\beta)$ no longer provides a good quantum number and Weyl tableaux are not good basis labels for interacting states\footnote{Also note that the first two equalities of (\ref{v3:abg}) are not equivalent under exchange of $\beta$ and $\gamma$ because in the choice of basis (\ref{basis:abg}), the state permutation subgroup generated by $\hat{U}(\alpha\beta)$ was diagonalized.}. To first order, the energy shifts for these levels are found by diagonalizing $\hat{V}^3$ in the $[21]$ sector to find
\begin{eqnarray}\label{21split}
v_\pm &=& v_{\pp{\alpha^2}\pp{\beta^2}} + v_{\pp{\beta^2}\pp{\gamma^2}}  + v_{\pp{\alpha^2}\pp{\gamma^2}} \nonumber\\
&& \pm \sqrt{(v_{\pp{\alpha\beta}^2})^2 - v_{\pp{\alpha\beta}^2} v_{\pp{\beta\gamma}^2} + (v_{\pp{\beta\gamma}^2})^2 -  v_{\pp{\beta\gamma}^2} v_{\pp{\alpha\gamma}^2} + (v_{\pp{\alpha\gamma}^2})^2 - v_{\pp{\alpha\beta}^2} v_{\pp{\alpha\gamma}^2} }.
\end{eqnarray}
The magnitudes of these shifts are intermediate between the shifts in the totally symmetric and totally antisymmetric sectors. The corresponding eigenvectors also depend on the two-body matrix elements, but the algebraic form of the eigenvalues and eigenvectors in terms of two-body matrix elements do not.

\begin{figure}
\centering
\includegraphics[width=.9\linewidth]{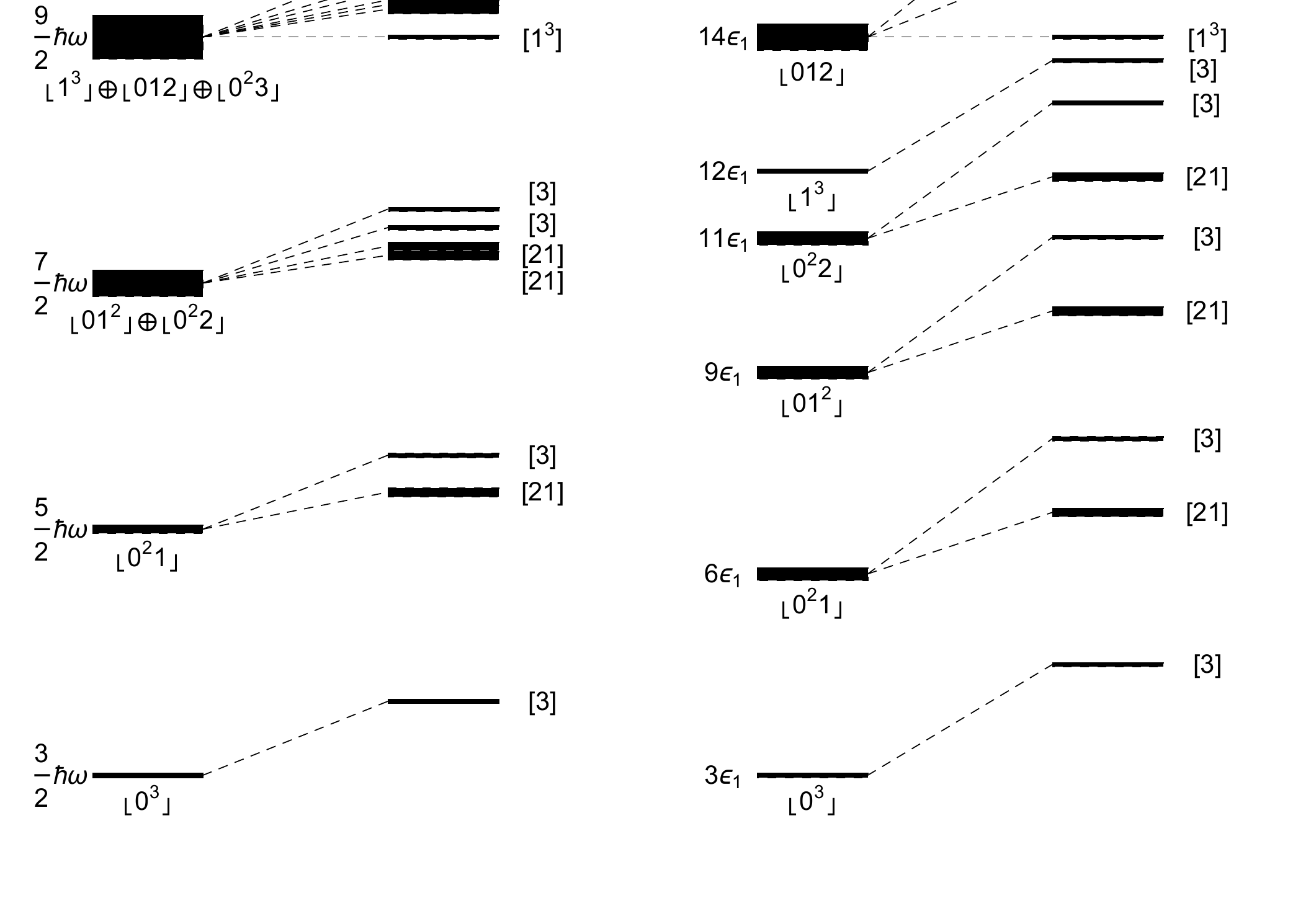}
\caption{This figure depicts the level splitting under weak contact interactions for a harmonic trap (left) and a hard wall trap (right). The energy scale for the two traps has been chosen so they have the same difference $\Delta E = E_\pp{012} - E_\pp{0^3}$ between the ground state with composition $\pp{0^3}$ and the energy level with composition $\pp{012}$. The strength of the contact interaction is $g = \Delta E/30$. The thickness of lines depicts the degeneracy of the energy level, or equivalently the dimension of the corresponding $\mathrm{S}_3$ irrep. For the highest non-interacting energy level, only the level corresponding to the unperturbed totally antisymmetric spatial state is depicted and the rest of the levels are cut off.}
\label{fig:weak3}
\end{figure}

To summarize, for weak interactions, a six-fold degenerate composition subspace like $\mathcal{K}^\pp{\alpha\beta\gamma}$ generally breaks into four levels. The biggest energy shift takes place for the totally symmetric state $\ykt{\tiny \young(\alpha\beta\gamma)}$ and the smallest energy shift for the totally antisymmetric state $\ykt{\tiny \young(\alpha,\beta,\gamma)}$. See Fig.~\ref{fig:weak3} for an example comparing the level splitting of the lowest few energy levels of a harmonic well and a hard wall well under weak contact interactions. In contrast, the states $\ykt{\tiny \young(\alpha\beta,\gamma) \, Y} $ and $\ykt{\tiny \young(\alpha\gamma,\beta) \,Y}$ mix under the interaction. Unlike the other composition subspaces $\mathcal{K}^\pp{\alpha^3}$ and $\mathcal{K}^\pp{\alpha^2\beta}$ or the totally symmetric and antisymmetric sectors of this composition subspace $\mathcal{K}^\pp{\alpha\beta\gamma}$, specific knowledge of the matrix elements is required to determine how the states with mixed symmetry split. The algebraic expression (\ref{21split}) for the level splitting of the two partially symmetric subspaces (which are relevant for bosons and fermions with $J\geq 2$) is universal, but only  in the weakest possible sense. In the sequel, it is hypothesized that for five particles and more, even this weakest kind of algebraic universality is broken because the diagonalizing the perturbation requires solving a quintic equation.

\subsection{Three Particles: Two-Body Contact Interaction at Unitarity}

\begin{figure}
\centering
\includegraphics[width=.3\linewidth]{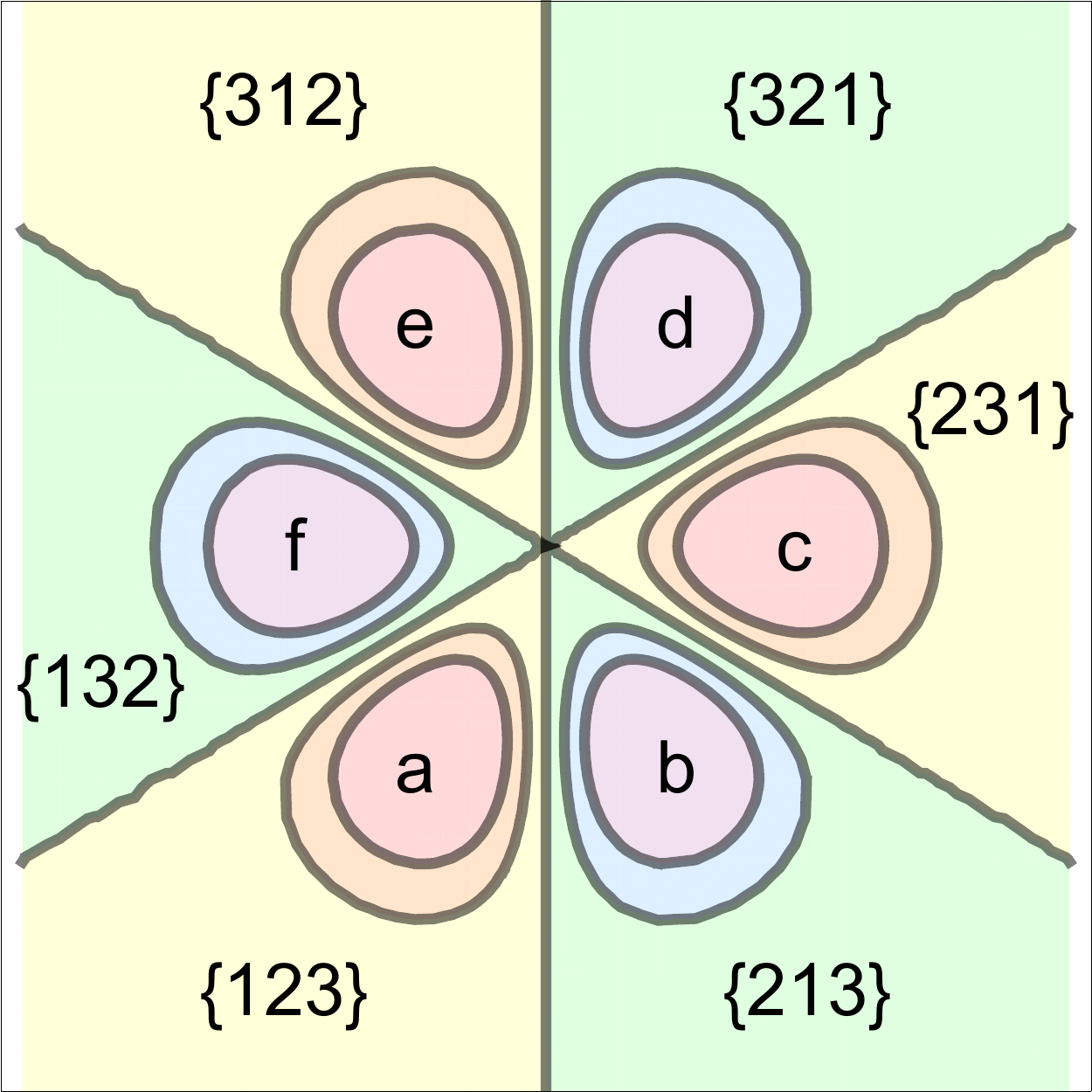}
\caption{This figure depicts a contour plot of the relative configuration space wave function for the totally antisymmetric lowest energy spatial state  of three particles in a harmonic well. The horizontal axis is the relative coordinate $r_1 = (q_1 - q_2)/\sqrt{2}$ and the vertical axis is the coordinate $r_2 = (q_1 + q_2 - 2 q_3)/\sqrt{6}$. The permutation $s$ in curly brackets denotes each sector $\mathcal{Q}_s$ where $q_{s_1} < q_{s_2} < q_{s_3}$.  The center-of-mass degree of freedom is perpendicular to this plane and not depicted. The sectors are also labeled by lower-case letters for convenience; see Table~\ref{tab:partvsorder}.}
\label{fig:3petal}
\end{figure}

To visualize the effect of contact interactions at the unitary limit $g \rightarrow \infty$, it is useful to visualize the coincidence manifold for three particles $\mathcal{V}^3$. This manifold is a structure in configuration space $\mathcal{Q}^3$ defined by the three planes $q_1=q_2$, $q_2=q_3$ and $q_1=q_3$. These three planes intersect at the line $q_1=q_2=q_3$ at angles of $2\pi/3$. The point symmetry of $\mathcal{V}^3$ is $\mathrm{D}_{6h}$, which is the same as $\mathrm{C}_3$ for the harmonic trap. See Fig.~\ref{fig:equip3} for a three-dimensional diagram of $\mathcal{V}^3$ and Fig.~\ref{fig:3petal} for a two-dimensional diagram of relative plane cross section of $\mathcal{V}^3$. 

The manifold $\mathcal{V}^3$ divides configuration space into six equivalent sectors, one for each order of particles $q_i<q_j < q_k$. Denote these sectors $\mathcal{Q}_s$, where $s$ is an element of  $\mathrm{S}_3$ expressed in permutation notation $s=\{s_1, s_2, s_3\}$ and the order is $q_{s_1}<q_{s_2} < q_{s_3}$. At the unitary limit, these sectors are dynamically isolated, but in the near-unitary limit they are connected by weak tunneling.

\subsection{Configuration Space Symmetry}

The configuration space representation $\underline{O}(p)$ of particle permutations $p\in\mathrm{P}_3$  map sectors $\mathcal{Q}_s$ onto each other like
\begin{equation}\label{sectrep}
\underline{O}(p)\mathcal{Q}_s = \mathcal{Q}_{sp^{-1}}.
\end{equation}
These maps are linear and continuous on $\mathcal{Q}^3$, mapping nearby points into nearby points in $\mathcal{Q}^3$, even if they are in different sectors.

The set $\mathrm{Q}_3$ of all possible sector exchange maps  is a group isomorphic to $\mathrm{S}_6$ and elements are represented as $6\times 6$ matrices $\underline{M}\in\mathrm{Q}_3$ with a single $1$ is each row and column and all the other matrix elements $0$. These act on the vector space of sectors like
\[
\underline{M} \mathcal{Q}_s = \mathcal{Q}_{s'}.
\]
Maps $\underline{M}$ that shuffle sectors are generally discontinuous transformations of $\mathcal{Q}^3$. The particle permutations $\underline{O}(p)$ are the only $6$ out of all $6!$ sector exchange maps that are linear and continuous on all of $\mathcal{Q}^3$.  

A particularly useful subset of sector exchange maps are ordering permutations $\mathfrak{O}_3$. Instead of permuting particle numbers wherever they appear in the sector label $s$, ordering permutations permute the order of numbers in $s$, no matter what particle numbers they are. For example, the ordering permutation $\underline{O}(AB)$ switches the order of the first and second number in the sector
\[
\underline{O}(AB) \mathcal{Q}_{\{312\}} = \mathcal{Q}_{\{132\}},
\]
whereas the particle permutation $\underline{O}(12)$ switches the positions of the numbers 1 and 2
\[
\underline{O}(12)\mathcal{Q}_{\{312\}} = \mathcal{Q}_{\{321\}}.
\]
See Table \ref{tab:partvsorder} for more examples of the difference between particle permutations and ordering permutations. The set of ordering permutations $\mathfrak{O}_3$ is isomorphic to $\mathrm{S}_3$ and acts `naturally' on the sectors, i.e.\ $\mathfrak{o}\in\mathfrak{O}_3$ 
\begin{equation}\label{ordsect}
\underline{O}(\mathfrak{o}) \mathcal{Q}_s = \mathcal{Q}_{\mathfrak{o} s}.
\end{equation}
Ordering transpositions $(AB)$ and $(BC)$ exchange sectors next to each other, i.e.\ the sectors most like to experience tunneling in the near-unitary limit.

\begin{table}[t]
\caption{This table compares the particle permutations $\underline{O}\{213\} \equiv \underline{O}(12)$ and $\underline{O}\{231\} \equiv \underline{O}(123)$ with the ordering permutations $\underline{O}\{BAC\} \equiv \underline{O}(AB)$ and $\underline{O}\{BCA\} \equiv \underline{O}(ABC)$. 
The sector notation is depicted by the lower-case letter in Fig.~\ref{fig:3petal}. Note that particle permutations have simple realizations in configuration space: two-cycles are reflections and three-cycles are rotations by $2\pi/3$. Ordering permutations do not have corresponding orthogonal transformations of configuration space, although they form an isomorphic group to $\mathrm{S}_3$ and map sectors onto sectors.}
\centering
\label{tab:partvsorder}
\begin{tabular}{|c|c||c|c|c|c|}
\hline
Sector & Label  & $ \underline{O}(12)$ &  $ \underline{O}(AB)$ & $ \underline{O}(123)$ &  $ \underline{O}(ABC)$ \\
\hline
$\{123\}$ & a & b & b & e & c\\
$\{213\}$ & b & a & a & f & f\\
$\{231\}$ & c & f & d & a & e\\
$\{321\}$ & d & e & c & b & b\\
$\{312\}$ & e & d & f & c & a\\
$\{132\}$ & f & c & e & d & d\\
\hline
\end{tabular}
\end{table}

\subsection{Kinematic Symmetry}

In the unitary limit, the six sectors $\mathcal{Q}_s$ are dynamically isolated. There is no tunneling between them and $\mathcal{V}^3$ is a nodal surface for every state with finite energy. More generally, because the sectors are dynamically isolated, the spatial Hilbert space of finite energy states $\KHS_\infty \subset \KHS$ can be decomposed into sectors
\begin{equation}\label{sectorp}
\KHS_\infty = \bigoplus_{s \in\mathrm{S}_3} \KHS_s
\end{equation}
where each $\KHS_s$ is the space of square integrable functions on $\mathcal{Q}_s$ that vanish on two half-planes from $\mathcal{V}^3$.

As for two particles, a basis for each sector is provided by the snippet basis. These are restrictions of the totally antisymmetric states like $\ykt{\tiny \young(\alpha,\beta,\gamma)}$ to each sector $\mathcal{Q}_s$:
\begin{equation}\label{bas:snip3}
\oybk{\bf q}{\tiny\young(\alpha,\beta,\gamma);s} = \left\{ \begin{array}{cc} \pi_s \sqrt{6} \oybk{\bf q}{\Yvcentermath1\tiny\young(\alpha,\beta,\gamma) \Yvcentermath0}  & \ \ \ \ {\bf q} \in\mathcal{Q}^3_p \\
0 & \ \ \ \ \mbox{else} \end{array} \right.,
\end{equation}
where $\pi_s$ is the sign of the permutation $s$: for two-cycles $\pi_s = -1$ and for three-cycles and the identity $\pi_s = 1$. This means that for every composition $\pp{\alpha\beta\gamma}$ of three distinct non-interacting energy eigenstates, there is a six-fold degenerate level with energy $E_\pp{\alpha\beta\gamma}=\epsilon_\alpha+\epsilon_\beta+\epsilon_\gamma$ in the spectrum $\sigma_3^\infty$. See Fig.~\ref{fig:3petal} for a depiction of the relative configuration space and lowest energy level $\ykt{\tiny \young(0,1,2)}$ for a harmonic trap.

Combining (\ref{sectrep}), (\ref{ordsect}), and (\ref{bas:snip3}), particle permutations $p\in\mathrm{P}_3$ act on the snippet basis like
\begin{equation}\label{partsnip}
\hat{U}(p) \ykt{\tiny\young(\alpha,\beta,\gamma);s} = \ykt{\tiny\young(\alpha,\beta,\gamma); s p^{-1}}
\end{equation}
and ordering permutations $\mathfrak{o}\in\mathfrak{O}_3$ act like
\begin{equation}\label{ordsnip}
\hat{U}(\mathfrak{o}) \ykt{\tiny\young(\alpha,\beta,\gamma);s} = \ykt{\tiny\young(\alpha,\beta,\gamma); \mathfrak{o} s }
\end{equation}
From this it is clear that all $p\in\mathrm{P}_3$ commute with all $\mathfrak{o}\in\mathfrak{O}_3$. Therefore, every  energy level $\KHS^\pp{\alpha\beta\gamma}$ carries a symmetry isomorphic to $\mathrm{S}_3 \times \mathrm{S}_3 \equiv \mathrm{S}_3^{\times 2}$. However, unlike state permutations which were a symmetry that depended on the compositions of the $\mathrm{K}_3^0$ irreps, the group $\mathfrak{O}_3$ is a symmetry of every finite-energy state in $\KHS_\infty$.

Denote the subspace spanned by the particular snippet basis (\ref{bas:snip3}) by $\KHS^\pp{\alpha\beta\gamma}_\infty$. This space carries the regular representation of $\mathrm{P}_3\sim\mathrm{S}_3$ isomorphic to $[3] \oplus 2 [21] \oplus  [1^3]$.
Just as the non-interacting irrep space $\mathcal{K}^\pp{\alpha\beta\gamma}$ carried an additional, independent copy  of $\mathrm{S}_3$ called state permutation symmetry $\mathfrak{P}_\pp{\alpha\beta\gamma}$ that  diagonalizes degenerate $\mathrm{P}_3$ irreps, ordering permutation symmetry $\mathfrak{O}_3$ can be used to diagonalize the degenerate $\mathrm{P}_3$ irreps in the snippet space $\KHS^\pp{\alpha\beta\gamma}_\infty$. 
Exploiting both particle permutation symmetry and ordering permutation symmetry, six symmetrized snippet vectors in $\ykt{\tiny\young(\alpha,\beta,\gamma);\mathfrak{Y}\,Y}\in \KHS^\pp{\alpha\beta\gamma}_\infty$ are denoted
\begin{equation}\label{bas:o3}
\ykt{\tiny\young(\alpha,\beta,\gamma);\tiny\young(ABC)}, \ykt{\tiny\young(\alpha,\beta,\gamma);\tiny\young(AC,B)\,\tiny\young(12,3)}, \ykt{\tiny\young(\alpha,\beta,\gamma);\tiny\young(AC,B)\,\tiny\young(13,2)},
\ykt{\tiny\young(\alpha,\beta,\gamma);\tiny\young(AB,C)\,\tiny\young(12,3)}, \ykt{\tiny\young(\alpha,\beta,\gamma);\tiny\young(AB,C)\,\tiny\young(13,2)}, \ykt{\tiny\young(\alpha,\beta,\gamma);\tiny\young(A,B,C)}.
\end{equation}
The Young tableaux $\mathfrak{Y}$ filled with $A$, $B$, and $C$ label the basis for irreps of the ordering permutation symmetry and the Young tableaux $Y$ filled with particle numbers label the basis for irreps of the particle permutation symmetry. The exact expression for these symmetrized vectors in terms of snippet bases vectors depends on the choice of subgroup chains for the complete set of commuting observables. For symmetric wells a good choice of observables is $\hat{C}^\pp{123}_2$, $\hat{C}^\pp{12}_2$, and $\hat{C}^\pp{AC}_2 = \hat{U}(AC)$ because within a snippet subspace with composition  $\pp{\alpha\beta\gamma}$ total parity inversion $\hat{\Pi}$ is proportional to $\hat{U}(AC)$
\begin{equation}
\hat{\Pi}\ykt{{\tiny\young(\alpha,\beta,\gamma)};\mathfrak{Y}\,Y} = -\pi_\alpha\pi_\beta\pi_\gamma \hat{U}(AC) \ykt{{\tiny\young(\alpha,\beta,\gamma)};\mathfrak{Y}\,Y}.
\end{equation}
With this convention, the states $\ykt{\tiny\young(\alpha,\beta,\gamma);\tiny\young(ABC)}$, $\ykt{\tiny\young(\alpha,\beta,\gamma);\tiny\young(AC,B)\,\tiny\young(12,3)}$, and $\ykt{\tiny\young(\alpha,\beta,\gamma);\tiny\young(AC,B)\,\tiny\young(13,2)}$ have the opposite parity from the totally antisymmetric state $\ykt{\tiny\young(\alpha,\beta,\gamma);\tiny\young(A,B,C)} = \ykt{\tiny\young(\alpha,\beta,\gamma)}$. See Fig.~8 for a depiction of the six states in $\KHS^\pp{\alpha\beta\gamma}_\infty$ using the complete set of commuting observables $\{ \hat{C}^\pp{123}_2, \hat{C}^\pp{12}_2, \hat{C}^\pp{AC}_2 \}$ to diagonalize the lowest energy level in the harmonic well.

\begin{figure}\label{fig:3unit}
\centering
\includegraphics[width=.7\linewidth]{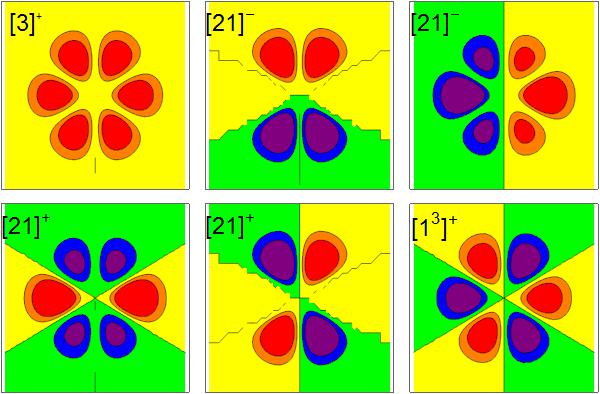}
\caption{This figure depicts a contour plot of the relative configuration space wave function for the lowest degenerate states in the unitary limit of the contact interaction for three particles in a harmonic well. The horizontal axis is the relative coordinate $r_1$ and the vertical axis is the coordinate $r_2 $. The states are depicted in the same order as (\ref{bas:o3}). }
\end{figure}

Note that one-component bosons can only populate the state $\ykt{\tiny\young(\alpha,\beta,\gamma);\tiny\young(ABC)}$ and one-component fermions can only populate the state $\ykt{\tiny\young(\alpha,\beta,\gamma);\tiny\young(A,B,C)}$. To include spin, consider  spin-1/2 distinguishable particles with $\SHS =\mathbb{C}^8$. Each energy level $\KHS^\pp{\alpha\beta\gamma}_\infty \times \SHS$  is 48-fold degenerate because there are eight spin states for every spatial state. Following the method described for the non-interacting, identical particles with spin in Section 4.2, using the Clebsch-Gordan series for $\mathrm{S}_3$ one can show that there are four two-component bosonic (fermionic) states with spatial symmetry $[3]$ ($[1^3]$) and total spin $s=3/2$ and four with spatial symmetry $[21]$ and total spin $s=1/2$.

\subsection{Near-Unitary Limit}

Another use for ordering permutation symmetry is looking at the near-unitary limit. In this limit there is tunneling between adjacent sectors. For an asymmetric well, the tunneling operator $\hat{T}$ can be parameterized
\begin{equation}\label{eq:tun3}
\hat{T} = -t \, \hat{U}(AB)  - u \, \hat{U}(BC) - (t+u) \hat{U}(e),
\end{equation}
where $t$ and $u$ are the tunneling amplitudes for the first and second particle and the second and third particle~\cite{Deuretzbacher2014}:
\begin{eqnarray}
t &=& \frac{6}{g} \int_{-\infty}^{+\infty} dq_3 \, \int_{-\infty}^{q_3} dq_2  \left|\frac{\partial \ybk{\bf q}{\tiny\young(\alpha,\beta,\gamma)}}{\partial q_1}\right|^2_{q_1=q_2}\nonumber\\
u &=& \frac{6}{g} \int_{-\infty}^{+\infty} dq_3 \, \int_{-\infty}^{q_3} dq_1  \left|\frac{\partial \ybk{\bf q}{\tiny\young(\alpha,\beta,\gamma)}}{\partial q_2}\right|^2_{q_2=q_3}.
\end{eqnarray}
The coefficients $t$ and $u$ depend on trap shape through the one-particle eigenstate wave functions, and for a symmetric trap $t=u$.  The third term in (\ref{eq:tun3}) which is  proportional to the identity renormalizes the energy shift such that the totally antisymmetric state undergoes no change in energy. The first order energy shifts are given by the eigenvalues of this matrix formed by the matrix elements $\ybr{\tiny\young(\alpha,\beta,\gamma); \mathfrak{Y}\,Y} \hat{T} \ykt{\tiny\young(\alpha,\beta,\gamma); \mathfrak{Y}'\,Y}$. The eigenvalues and degeneracies are
\begin{eqnarray}
\Delta E_{[3]} &=& -2t -2u\ \mbox{(singly-degenerate)},\nonumber\\
\Delta E_{[21]} &=& -t -u - \sqrt{t^2 -tu + u^2}\ \mbox{(doubly-degenerate)},\nonumber\\
\Delta E_{[21]'} &=& -t -u + \sqrt{t^2 -tu + u^2}\ \mbox{(doubly-degenerate)},\nonumber\\
\Delta E_{[1^3]} &=& 0 \ \mbox{(singly-degenerate)}.
\end{eqnarray}
The first and last eigenvalues correspond to symmetric and antisymmetric states, and the middle two eigenvalues are for two different two-dimensional mixed-symmetry irreps. Since the ratio of $t$ to $u$ depends on trap shape, the first-order energy depends on trap shape for asymmetric traps. The symmetric $[3]$ and antisymmetric states $[1^3]$ are universal for any trap, but energy eigenstates with mixed symmetry $[21]$ are not, although there are universal algebraic expressions in terms of $t$ and $u$.

However, for symmetric traps $t=u$ and the eigenvalues become
\begin{eqnarray}
\Delta E_{[3]^\pm} &=& -4t\ \mbox{(singly-degenerate)},\nonumber\\
\Delta E_{[21]^\mp} &=& -3t \ \mbox{(doubly-degenerate)},\nonumber\\
\Delta E_{[21]^\pm} &=& -t\ \mbox{(doubly-degenerate)},\nonumber\\
\Delta E_{[1^3]^\mp} &=& 0 \ \mbox{(singly-degenerate)}.
\end{eqnarray}
With the addition of parity symmetry, we find that although the eigenvalues depend on the trap shape through $t$, the states do not. The ordering permutation observable $\hat{U}(AC)\sim\hat{\Pi}$ is sufficient to distinguish the two copies of the $[21]$ irreps. The energy shifts $-3t$ and $0$ are for states with the same parity $\pi_\alpha\pi_\beta\pi_\gamma$ as the original fermionic state $\ykt{\tiny\young(\alpha,\beta,\gamma)}$ and the energy shifts $-4t$ and $-t$ are states with opposite parity $-\pi_\alpha\pi_\beta\pi_\gamma$. These results agree with recent work on the near unitary limit for harmonic traps \cite{Deuretzbacher2014, Volosniev2013, Levinsen2014, Gharashi2015} where $t=u= 3^3/(2^3\sqrt{2\pi} g )$.

\section{Conclusion of Part I}

What lessons have been learned from this analysis of the symmetries for one, two and three trapped particles? What methods can be extended to more particles, and what methods lose utility?

One set of observations is about how symmetries of the individual particles build into symmetries of the total system. Any system with a Hamiltonian that is a sum of identical sub-Hamiltonians, each with symmetry $\mathrm{G}_1$, must have at least the minimal the symmetry structure $\mathrm{P}_N \ltimes (\mathrm{G}_1)^{\times N}$. In particular, the configuration space symmetry and the kinematic symmetries of two or three non-interacting particles satisfy
\begin{equation}\label{eq:nonint}
\mathrm{C}_N^0 \supseteq \mathrm{P}_N \ltimes \mathrm{C}_1^{\times N}\ \mbox{and}\ \mathrm{K}_N^0  \supseteq \mathrm{P}_N \ltimes \mathrm{K}_1^{\times N}.
\end{equation}
When these groups are greater than the minimal symmetry, then that signals the emergence of a true multiparticle symmetry. For two and three particle systems, the groups $\mathrm{C}_2^0$ and $\mathrm{C}_3^0$ are finite order point groups in two and three dimensions, realized by orthogonal transformations of configuration space. Unless there are emergent symmetries or accidental degeneracies, the irreps of $\mathrm{K}_2^0$ and $\mathrm{K}_3^0$ are sufficient to explain the degeneracy pattern of the non-interacting energies. The main difference between the case of two and three particles is that for three particles there can be multiple irreps of the symmetric group $\mathrm{S}_3$ with the same energy, even without emergent or accidental symmetries. State permutation symmetry is introduced to build observables that distinguish these degenerate spaces.  All of this is extended to more particles in the sequel article using the mathematics of permutation modules, and the minimal construction (\ref{eq:nonint}) is shown to be algebraically universal, i.e.\ a complete set of observables that is independent of trap shape can always be built out of single particle operators.

For systems with Galilean-invariant, spin-independent two-body interactions, then the symmetry (\ref{eq:nonint}) is broken and algebraic universality is lost. However, the symmetry that remains is always at least 
\begin{equation}\label{eq:int}
\mathrm{C}_N \supseteq \mathrm{P}_N \times \mathrm{C}_1\ \mbox{and}\ \mathrm{K}_N \supseteq \mathrm{P}_N \times \mathrm{K}_1.
\end{equation}
The configuration space symmetries again have geometrical realizations as point groups that preserve both the trap equipotentials and the coincidence manifold. The irreps of the kinematic symmetries are irreps of $\mathrm{S}_N$ with additional quantum numbers inherited from the single-particle trap, e.g.\ parity for symmetric traps and center-of-mass quantum number for harmonic traps. The energy eigenvalues and eigenstates are certainly not universal, but the symmetries can be used to decrease the computational scale of exact diagonalization or perturbation theory by using basis vectors from the non-interacting spatial Hilbert space reduced into irreps of $\mathrm{K}_N$.

For weak interactions, two and three particles have different degrees of algebraic universality. The two-body matrix elements, which depend on the trap and the interaction, are required in order to find the specific energy shift at first order. However, the splitting pattern and the energy eigenstates are independent of interaction and trap shape for two particles and are therefore algebraically universal in a strong sense because the two-body matrix elements are not required. For three particles, there is less universality under weak perturbations. Some splitting features remain independent of trap and interaction, but the full solution for some first-order eigenstates requires algebraic expressions involving the two-body matrix elements. In the sequel article even this very limited universal feature is hypothesized to fail for $N \geq 5$.

The contact interaction restores some algebraic universality, and this can be partially understood as a manifestation of the state permutation symmetry of the two-body matrix elements. Also, for the contact interaction the unitary limit is sensible because the delta interaction enforces nodal surfaces in configuration space. These surfaces prevent particles tunneling past each other, and break configuration space into sectors with a particular order. The emergent combination of ordering permutation symmetry and particle permutation symmetry provides enough structure for complete algebraic solution of the Hamiltonian in the algebraic limit assuming knowledge of the single particle spectrum. The near unitary mapping is also algebraically universal for two particles, but for three particles in an asymmetric trap information about the trap shape (or equivalently the single particle wave functions) is required to solve for the level splitting. For symmetric traps, the extra information provided by parity is enough to restore algebraic universality.

\begin{acknowledgements}
Many colleagues provided me insight and help with this research, but special thanks goes to the organizers and participants of the workshop ``Universality in Few-Body Systems: Theoretical Challenges and New Directions'' at the Institute for Nuclear Theory at the University of Washington where some of this work was done and much of it discussed in Spring 2014. Special thanks goes to D.~Blume, K.~Daily, N.~Mehta, S.~Tan, and A.~Volosniev. I am also sincerely grateful to a referee whose comments improved the article and (I hope) made it more useful to a broader audience.
\end{acknowledgements}

\end{document}